\pdfoutput=1
\documentclass[12pt,a4paper]{article}
\usepackage{ifthen} % for conditional statements
\newboolean{pdflatex}
\setboolean{pdflatex}{true} % False for eps figures 

\newboolean{articletitles}
\setboolean{articletitles}{true} % False removes titles in references

\newboolean{uprightparticles}
\setboolean{uprightparticles}{false} %True for upright particle symbols

\def\paperauthors{LHCb collaboration} % Leave as is for PAPER, CONF and FIGURE
\def\paperasciititle{Measurement of the W boson mass} % Set ASCII title here !! MAKE sure it's only ASCII characters !! 
\def\papertitle{Measurement of the \PW boson mass}
\def\paperkeywords{{High Energy Physics}, {LHCb}} % Comma separated list
\def\papercopyright{\the\year\ CERN for the benefit of the LHCb collaboration} % new since 9/Apr/2018
\def\paperlicence{CC BY 4.0 licence}
\def\paperlicenceurl{https://creativecommons.org/licenses/by/4.0/}

\usepackage[top=1in, bottom=1.25in, left=1in, right=1in]{geometry}

\usepackage{microtype}
\usepackage{lineno}  % for line numbering during review
\usepackage{xspace} % To avoid problems with missing or double spaces after
\usepackage{caption} %these three command get the figure and table captions automatically small

\usepackage{graphicx}  % to include figures (can also use other packages)
\usepackage{color}
\usepackage{colortbl}
\graphicspath{{./figs/}} % Make Latex search fig subdir for figures
\usepackage{amsmath} % Adds a large collection of math symbols
\usepackage{amssymb}
\usepackage{amsfonts}
\usepackage{upgreek} % Adds in support for greek letters in roman typeset

\newcommand*\patchAmsMathEnvironmentForLineno[1]{%
\expandafter\let\csname old#1\expandafter\endcsname\csname #1\endcsname
\expandafter\let\csname oldend#1\expandafter\endcsname\csname
end#1\endcsname
 \renewenvironment{#1}%
   {\linenomath\csname old#1\endcsname}%
   {\csname oldend#1\endcsname\endlinenomath}%
}
\newcommand*\patchBothAmsMathEnvironmentsForLineno[1]{%
  \patchAmsMathEnvironmentForLineno{#1}%
  \patchAmsMathEnvironmentForLineno{#1*}%
}
\AtBeginDocument{%
\patchBothAmsMathEnvironmentsForLineno{equation}%
\patchBothAmsMathEnvironmentsForLineno{align}%
\patchBothAmsMathEnvironmentsForLineno{flalign}%
\patchBothAmsMathEnvironmentsForLineno{alignat}%
\patchBothAmsMathEnvironmentsForLineno{gather}%
\patchBothAmsMathEnvironmentsForLineno{multline}%
\patchBothAmsMathEnvironmentsForLineno{eqnarray}%
}

\usepackage{hyperxmp}

\usepackage[pdftex,
            pdfauthor={\paperauthors},
            pdftitle={\paperasciititle},
            pdfkeywords={\paperkeywords},
            pdfcopyright={Copyright (C) \papercopyright},
            pdflicenseurl={\paperlicenceurl}]{hyperref}
\usepackage[colorinlistoftodos,textsize=scriptsize]{todonotes}

\usepackage[bottom,flushmargin,hang,multiple]{footmisc}

\usepackage[all]{hypcap} % Internal hyperlinks to floats.

\usepackage{xspace} 
\usepackage{upgreek}

\def\lhcb   {\mbox{LHCb}\xspace}

\def\MagUp {\mbox{\em Mag\kern -0.05em Up}\xspace}

\ifthenelse{\boolean{uprightparticles}}%
{

 \def\Pmu         {\ensuremath{\upmu}\xspace}                 
 \def\Pnu         {\ensuremath{\upnu}\xspace}

 \def\Ppsi        {\ensuremath{\uppsi}\xspace}

 \def\PDelta      {\ensuremath{\Delta}\xspace}                 
 \def\PXi         {\ensuremath{\Xi}\xspace}                 
 \def\PLambda     {\ensuremath{\Lambda}\xspace}                 
 \def\PSigma      {\ensuremath{\Sigma}\xspace}                 
 \def\POmega      {\ensuremath{\Omega}\xspace}                 
 \def\PUpsilon    {\ensuremath{\Upsilon}\xspace}

 \def\PB      {\ensuremath{\mathrm{B}}\xspace}                 
                  
 \def\PD      {\ensuremath{\mathrm{D}}\xspace}

 \def\PJ      {\ensuremath{\mathrm{J}}\xspace}                 
 \def\PK      {\ensuremath{\mathrm{K}}\xspace}

 \def\PW      {\ensuremath{\mathrm{W}}\xspace}

 \def\PZ      {\ensuremath{\mathrm{Z}}\xspace}                 
                  
 \def\Pb      {\ensuremath{\mathrm{b}}\xspace}                 
 \def\Pc      {\ensuremath{\mathrm{c}}\xspace}

 \def\Pi      {\ensuremath{\mathrm{i}}\xspace}

 \def\Ps      {\ensuremath{\mathrm{s}}\xspace}

 \def\thebaroffset{0.0em}
}
{

 \def\Pmu         {\ensuremath{\mu}\xspace}                 
 \def\Pnu         {\ensuremath{\nu}\xspace}

 \def\Ppsi        {\ensuremath{\psi}\xspace}                 
                  
 \mathchardef\PDelta="7101
 \mathchardef\PXi="7104
 \mathchardef\PLambda="7103
 \mathchardef\PSigma="7106
 \mathchardef\POmega="710A
 \mathchardef\PUpsilon="7107
                  
 \def\PB      {\ensuremath{B}\xspace}                 
                  
 \def\PD      {\ensuremath{D}\xspace}

 \def\PJ      {\ensuremath{J}\xspace}                 
 \def\PK      {\ensuremath{K}\xspace}

 \def\PW      {\ensuremath{W}\xspace}  
 \def\PWp      {\ensuremath{W^+}\xspace}  
 \def\PWm      {\ensuremath{W^-}\xspace}

 \def\PZ      {\ensuremath{Z}\xspace}                 
                  
 \def\Pb      {\ensuremath{b}\xspace}                 
 \def\Pc      {\ensuremath{c}\xspace}

 \def\Pi      {\ensuremath{i}\xspace}

 \def\Ps      {\ensuremath{s}\xspace}

 \def\thebaroffset{0.18em}
}
\newcommand{\offsetoverline}[2][\thebaroffset]{\kern #1\overline{\kern -#1 #2}}%

\makeatletter
\ifcase \@ptsize \relax% 10pt
  \newcommand{\miniscule}{\@setfontsize\miniscule{4}{5}}% \tiny: 5/6
\or% 11pt
  \newcommand{\miniscule}{\@setfontsize\miniscule{5}{6}}% \tiny: 6/7
\or% 12pt
  \newcommand{\miniscule}{\@setfontsize\miniscule{5}{6}}% \tiny: 6/7
\fi
\makeatother

\DeclareRobustCommand{\optbar}[1]{\shortstack{{\miniscule (\rule[.5ex]{1.25em}{.18mm})}
  \\ [-.7ex] $#1$}}

\def\muon       {{\ensuremath{\Pmu}}\xspace}

\def\neu        {{\ensuremath{\Pnu}}\xspace}

\def\Z      {{\ensuremath{\PZ}}\xspace}

\def\squark    {{\ensuremath{\Ps}}\xspace}

\def\cquark    {{\ensuremath{\Pc}}\xspace}

\def\bquark    {{\ensuremath{\Pb}}\xspace}

\def\KorKbar {\kern \thebaroffset\optbar{\kern -\thebaroffset \PK}{}\xspace}

\def\D       {{\ensuremath{\PD}}\xspace}

\def\DorDbar {\kern \thebaroffset\optbar{\kern -\thebaroffset \PD}\xspace}

\def\Dp      {{\ensuremath{\D^+}}\xspace}
\def\Dm      {{\ensuremath{\D^-}}\xspace}

\def\DpDm    {\ensuremath{\Dp {\kern -0.16em \Dm}}\xspace}

\def\B       {{\ensuremath{\PB}}\xspace}

\def\BorBbar {\kern \thebaroffset\optbar{\kern -\thebaroffset \PB}\xspace}

\def\Bd      {{\ensuremath{\B^0}}\xspace}

\def\BdorBdbar {\kern \thebaroffset\optbar{\kern -\thebaroffset \Bd}\xspace}

\def\Bs      {{\ensuremath{\B^0_\squark}}\xspace}

\def\BsorBsbar {\kern \thebaroffset\optbar{\kern -\thebaroffset \Bs}\xspace}

\def\jpsi     {{\ensuremath{{\PJ\mskip -3mu/\mskip -2mu\Ppsi}}}\xspace}

\def\Upsilonres  {{\ensuremath{\PUpsilon}}\xspace}
\def\Y#1S{\ensuremath{\PUpsilon{(#1S)}}\xspace}
\def\OneS  {{\Y1S}\xspace}

\def\LorLbar     {\kern \thebaroffset\optbar{\kern -\thebaroffset \PLambda}\xspace}

\def\to                 {\ensuremath{\rightarrow}\xspace}

\def\AT#1     {\ensuremath{A_{\mathrm{T}}^{#1}}\xspace}           % 2

\def\C#1      {\ensuremath{\mathcal{C}_{#1}}\xspace}                       % 9
\def\Cp#1     {\ensuremath{\mathcal{C}_{#1}^{'}}\xspace}                    % 7
\def\Ceff#1   {\ensuremath{\mathcal{C}_{#1}^{\mathrm{(eff)}}}\xspace}        % 9  
\def\Cpeff#1  {\ensuremath{\mathcal{C}_{#1}^{'\mathrm{(eff)}}}\xspace}       % 7
\def\Ope#1    {\ensuremath{\mathcal{O}_{#1}}\xspace}                       % 2
\def\Opep#1   {\ensuremath{\mathcal{O}_{#1}^{'}}\xspace}                    % 7

\newcommand{\nospaceunit}[1]{\ensuremath{\text{#1}}}       
\newcommand{\aunit}[1]{\ensuremath{\text{\,#1}}}       
\newcommand{\tev}{\aunit{Te\kern -0.1em V}\xspace}
\newcommand{\gev}{\aunit{Ge\kern -0.1em V}\xspace}
\newcommand{\mev}{\aunit{Me\kern -0.1em V}\xspace}
\newcommand{\kev}{\aunit{ke\kern -0.1em V}\xspace}
\newcommand{\ev}{\aunit{e\kern -0.1em V}\xspace}
 
\newcommand{\mevc}{\ensuremath{\aunit{Me\kern -0.1em V\!/}c}\xspace}
\newcommand{\gevc}{\ensuremath{\aunit{Ge\kern -0.1em V\!/}c}\xspace}
\newcommand{\mevcc}{\ensuremath{\aunit{Me\kern -0.1em V\!/}c^2}\xspace}
\newcommand{\gevcc}{\ensuremath{\aunit{Ge\kern -0.1em V\!/}c^2}\xspace}
\def\mum  {\ensuremath{\,\upmu\nospaceunit{m}}\xspace}

\def\fb   {\ensuremath{\aunit{fb}}\xspace}
\def\invfb   {\ensuremath{\fb^{-1}}\xspace}

\newcommand{\chisqip}{\ensuremath{\chi^2_{\text{IP}}}\xspace}

\def\deriv {\ensuremath{\mathrm{d}}}

\def\gsim{{~\raise.15em\hbox{$>$}\kern-.85em
          \lower.35em\hbox{$\sim$}~}\xspace}
\def\lsim{{~\raise.15em\hbox{$<$}\kern-.85em
          \lower.35em\hbox{$\sim$}~}\xspace}

\def\pt         {\ensuremath{p_{\mathrm{T}}}\xspace}

\def\ptot       {\ensuremath{p}\xspace}

\def\geant      {\mbox{\textsc{Geant4}}\xspace}

\def\herwig     {\mbox{\textsc{Herwig}}\xspace}

\def\photos     {\mbox{\textsc{Photos}}\xspace}
\def\powheg     {\mbox{\textsc{Powheg}}\xspace}
\def\pythia     {\mbox{\textsc{Pythia}}\xspace}
\def\dyturbo     {\mbox{\textsc{DYTurbo}}\xspace}

\def\powhegbox     {\mbox{\textsc{POWHEGBoxV2}}\xspace}
\def\powhegpythia     {\mbox{\textsc{POWHEGPythia}}\xspace}
\def\powhegherwig     {\mbox{\textsc{POWHEGHerwig}}\xspace}
\def\herwignlo     {\mbox{\textsc{Herwig}}\xspace}

\def\tell1  {TELL1\xspace}
\def\ukl1   {UKL1\xspace}

\newcommand{\eg}{\mbox{\itshape e.g.}\xspace}

\usepackage{cite} % Allows for ranges in citations
\usepackage{mciteplus}
\xspaceaddexceptions{]\}}
\def\DataMWValue {80354}
\def\DataMWTot {32}
\def\DataMWStat {23}
\def\DataMWExp {10}
\def\DataMWTh {17}
\def\DataMWPDF {9}

\def\NNPDFmWval {80362}
\def\MSHTmWval {80351}
\def\CTmWval {80350}
\def\NNPDFmWerr {9}
\def\MSHTmWerr {7}
\def\CTmWerr {12}
\def\phiCS  {{\ensuremath{\varphi}}\xspace}
\def\thetaCS  {{\ensuremath{\vartheta}}\xspace}
\def\chisqip  {{\ensuremath{\chi^2_{\rm IP}}}\xspace}

\def\Zmm     {{\ensuremath{\Z \to \mu\mu}}\xspace}

\def\IKT    {{\ensuremath{k_{\rm T}^{\rm intr}}}\xspace}

\def\Upsmumu  {{\ensuremath{\OneS \to \mu\mu}}\xspace}

\def\DefaultFigWidth{0.49\textwidth}
\usepackage{longtable} % only for template; not usually to be used in PAPERs

\begin{document}

%%%%%%%%%%%%%%%%%%%%%%%%%
%%%%% Title     %%%%%%%%%
%%%%%%%%%%%%%%%%%%%%%%%%%
\renewcommand{\thefootnote}{\fnsymbol{footnote}}
\setcounter{footnote}{1}

% %%%%%%% CHOOSE TITLE PAGE--------
%\onecolumn
%\input{title-LHCb-INT}
%\input{title-LHCb-ANA}
%\input{title-LHCb-CONF}
%\input{title-LHCb-FIGURE}
% ===============================================================================
% Purpose: LHCb-PAPER journal paper title page template
% Author: 
% Created on: 2010-09-25
% ===============================================================================

%%%%%%%%%%%%%%%%%%%%%%%%%
%%%%%  TITLE PAGE  %%%%%%
%%%%%%%%%%%%%%%%%%%%%%%%%
\begin{titlepage}
\pagenumbering{roman}

% Header ---------------------------------------------------
\vspace*{-1.5cm}
\centerline{\large EUROPEAN ORGANIZATION FOR NUCLEAR RESEARCH (CERN)}
\vspace*{1.5cm}
\noindent
\begin{tabular*}{\linewidth}{lc@{\extracolsep{\fill}}r@{\extracolsep{0pt}}}
\ifthenelse{\boolean{pdflatex}}% Logo format choice
{\vspace*{-1.5cm}\mbox{\!\!\!\includegraphics[width=.14\textwidth]{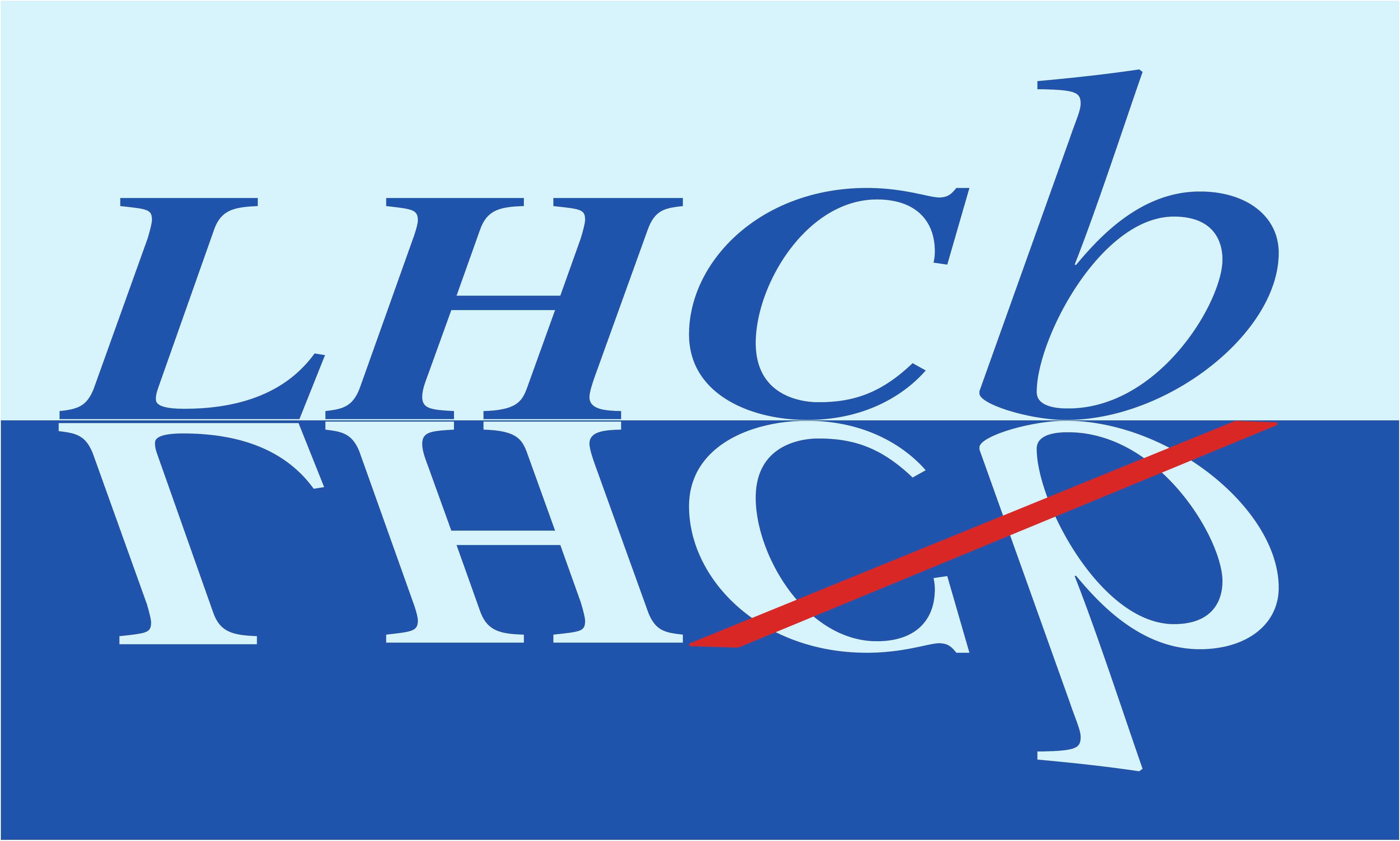}} & &}%
{\vspace*{-1.2cm}\mbox{\!\!\!\includegraphics[width=.12\textwidth]{lhcb-logo.eps}} & &}%
\\
 & & CERN-EP-2021-170 \\  % ID 
 & & LHCb-PAPER-2021-024 \\  % ID 
 & & \today \\ % Date - Can also hardwire e.g.: 23 March 2010
 & & \\
% not in paper \hline
\end{tabular*}

%\vspace*{4.0cm}
\vspace*{2.0cm}

% Title --------------------------------------------------
{\normalfont\bfseries\boldmath\huge
\begin{center}
% DO NOT EDIT HERE. Instead edit macro in main.tex to keep metadata correct
  \papertitle 
\end{center}
}

%\vspace*{2.0cm}
\vspace*{1.5cm}

% Authors -------------------------------------------------
\begin{center}
%In the footnote, replace 'paper' by 'Letter' in case of submission to PRL or PLB 
% Edit macro in main.tex to keep metadata correct
\paperauthors\footnote{Authors are listed at the end of this paper.}
\end{center}

\vspace{\fill}

% Abstract -----------------------------------------------
\begin{abstract}
  \noindent
  The \PW boson mass is measured using proton-proton collision data at $\sqrt{s}=13\tev$ corresponding to an integrated luminosity of 1.7\invfb recorded during 2016 by the LHCb experiment.
  With a simultaneous fit of the muon $q/\pt$ distribution of a sample of $\PW \to \muon\neu$ decays
  and the $\phi^*$ distribution of a sample of $\PZ\to\muon\muon$ decays the \PW boson mass is determined to be 
  \begin{equation*}
    m_{W} = \text{\DataMWValue} \pm \DataMWStat_{\rm stat} \pm \DataMWExp_{\rm exp} \pm \DataMWTh_{\rm theory} \pm \DataMWPDF_{\rm PDF}~\mathrm{MeV},
  \end{equation*}
 where uncertainties correspond to contributions from statistical, experimental systematic, theoretical and parton distribution function sources.
 This is an average of results based on three recent global parton distribution function sets. The measurement agrees well with the prediction of the global electroweak fit and with previous measurements.
\end{abstract}

\vspace*{2.0cm}

\begin{center}
Published in JHEP 01 (2022) 036.
%  Submitted to
%  JHEP
\end{center}

\vspace{\fill}

{\footnotesize 
% Edit macro in main.tex to keep metadata correct
\centerline{\copyright~\papercopyright. \href{\paperlicenceurl}{\paperlicence}.}}
\vspace*{2mm}

\end{titlepage}

%%%%%%%%%%%%%%%%%%%%%%%%%%%%%%%%
%%%%%  EOD OF TITLE PAGE  %%%%%%
%%%%%%%%%%%%%%%%%%%%%%%%%%%%%%%%

%  empty page follows the title page ----
\newpage
\setcounter{page}{2}
\mbox{~}
%\newpage
%
%% Author List ----------------------------
%%  You need to get a new author list!
%\input{LHCb_authorlist.tex}
%
%The author list for journal publications is provided by the Membership Committee shortly after 'approval to go to paper' has been given.
%%It will be made available on the page
%%\verb!http://www.physik.uzh.ch/~strauman/forMemCo/LHCb-PAPER-XXXX-XXX/! .
%It will be sent to you by email shortly after a paper number has beens assigned.
%The author list should be included already at first circulation, 
%to allow new members of the collaboration to verify whether they have been included correctly.
%Occasionally a misspelled name is corrected or associated institutions become full members.
%In that case, a new author list will be sent to you.
%In case line numbering doesn't work well after including the authorlist, try moving the \verb!\bigskip! after the last author to a separate line.
%
%
%The authorship for Conference Reports should be ``The LHCb
%  collaboration'', with a footnote giving the name(s) of the contact
%  author(s), but without the full list of collaboration names.

%\twocolumn
% %%%%%%%%%%%%% ---------

\renewcommand{\thefootnote}{\arabic{footnote}}
\setcounter{footnote}{0}

%%%%%%%%%%%%%%%%%%%%%%%%%%%%%%%%
%%%%%  Table of Content   %%%%%%
%%%%%%%%%%%%%%%%%%%%%%%%%%%%%%%%
%%%% Uncomment if desired
%\tableofcontents
\cleardoublepage

%%%%%%%%%%%%%%%%%%%%%%%%%
%%%%% Main text %%%%%%%%%
%%%%%%%%%%%%%%%%%%%%%%%%%

\pagestyle{plain} % restore page numbers for the main text
\setcounter{page}{1}
\pagenumbering{arabic}

%% Uncomment during review phase. 
%% Comment before a final submission.
%\linenumbers

%% This is the main body
%% It is useful to have a single file so comemnts are not missed in overleaf.
\section{Introduction}

The \PW boson mass ($m_W$) is directly related to electroweak (EW) symmetry breaking in the Standard Model (SM)~\cite{SM1,SM2,SM3}.
At tree level, $m_W = g v /2$ 
where $g$ is the weak-isospin coupling and $v$ is the vacuum expectation value of the Higgs field.
Going beyond tree level the boson masses and couplings receive loop corrections.
The value of $m_W$ is related to the precisely measured fine-structure constant ($\alpha$), the mass of the \PZ boson
($m_Z$) and the Fermi constant ($G_F$), as~\cite{Awramik:2003rn,PhysRevD.22.971}
\begin{equation}
\label{eq:mWfromPEW}
m_W^2 \left( 1 - \frac{m_W^2}{m_Z^2} \right) = \frac{\pi \alpha}{\sqrt{2}G_F}( 1 + \Delta ),
\end{equation}
where $\Delta$ encapsulates the loop-level corrections.
A global fit of EW observables, excluding direct measurements of $m_W$, yields a prediction
of $m_W = 80354 \pm 7\mev$~\cite{Gfitter}.\footnote{Throughout this paper natural units with $c=1$ are used.}
This can be compared with direct measurements to test for possible beyond SM contributions to $\Delta$ in Eq.~\ref{eq:mWfromPEW}.
The 2020 PDG average of direct measurements 
is $m_W = 80379 \pm12\mev$~\cite{PDG2020}.
The sensitivity of the global EW fit to physics beyond the SM is primarily limited by the precision of the direct measurements of $m_W$~\cite{Gfitter}.
Furthermore, the uncertainty in the prediction is expected to reduce as the top-quark mass, which is the leading source of the parametric uncertainty,
is determined more precisely in the future.

The value of $m_W$ was measured to a precision of 33\mev at the Large Electron-Positron (LEP) collider~\cite{LEP:ew:2013} at CERN and to a precision of 16\mev in an average~\cite{Tevatron} of measurements by the CDF~\cite{CDF} and D0~\cite{D0} experiments at the Fermilab Tevatron collider.
The first measurement at the LHC was performed by the ATLAS collaboration and has an uncertainty of 19\mev~\cite{ATLAS}.
The hadron collider measurements are based on three observables in leptonic \PW boson decays, namely the transverse mass, missing transverse momentum and charged lepton transverse momentum ($\pt$).
At hadron colliders, the lepton $\pt$ is measured with good resolution but it is strongly influenced by the \PW boson transverse momentum distribution,
the modelling of which is a potential source of a limiting systematic uncertainty.
However, the resolution of the transverse mass is degraded by the pile-up of proton-proton interactions in the same bunch crossing.
Therefore, the lepton \pt was the most sensitive observable in the recent measurement performed by the ATLAS collaboration.
Despite being based on a small subset of the data recorded to date, the ATLAS measurement of $m_W$ is already limited by uncertainties in modelling \PW boson production,
in particular the parton distribution functions (PDFs) of the proton.

The potential for a measurement based on the muon $\pt$ with the LHCb experiment is studied in Ref.~\cite{Vesterinen}
It was estimated that LHCb data collected in LHC Run~2, at a proton-proton ($pp$) centre of mass energy $\sqrt{s}=13\tev$, would allow a measurement with a statistical precision of around 10\mev.
Owing to the complementary pseudorapidity ($\eta$) coverage of the LHCb experiment with respect to the  ATLAS and CMS experiments, it was demonstrated in Ref.~\cite{Vesterinen} that the PDF uncertainty could partially cancel in an average of $m_W$ measurements by the LHC experiments.

\begin{figure}[!b]\centering
\includegraphics[width=\DefaultFigWidth]{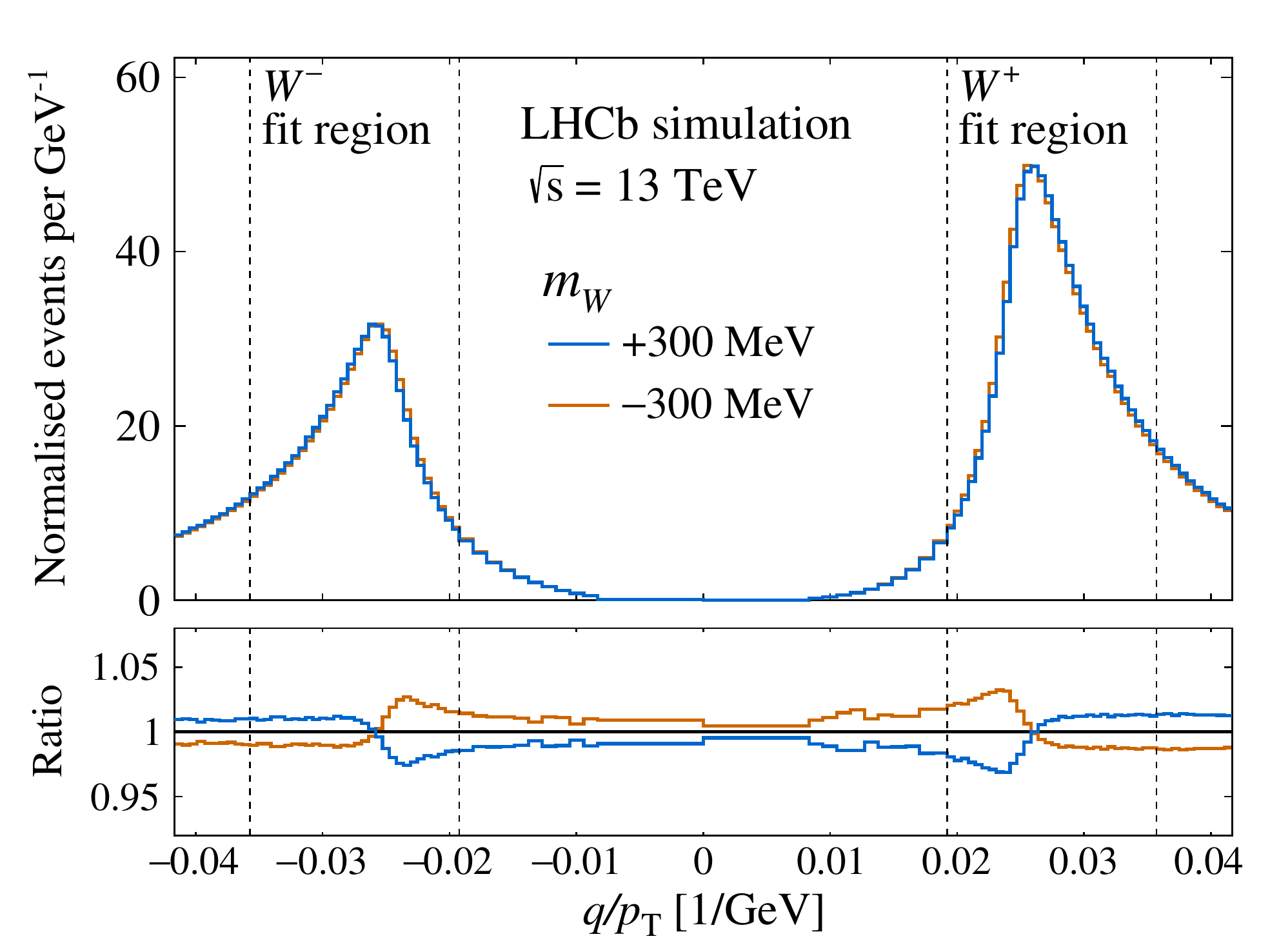}
\includegraphics[width=\DefaultFigWidth]{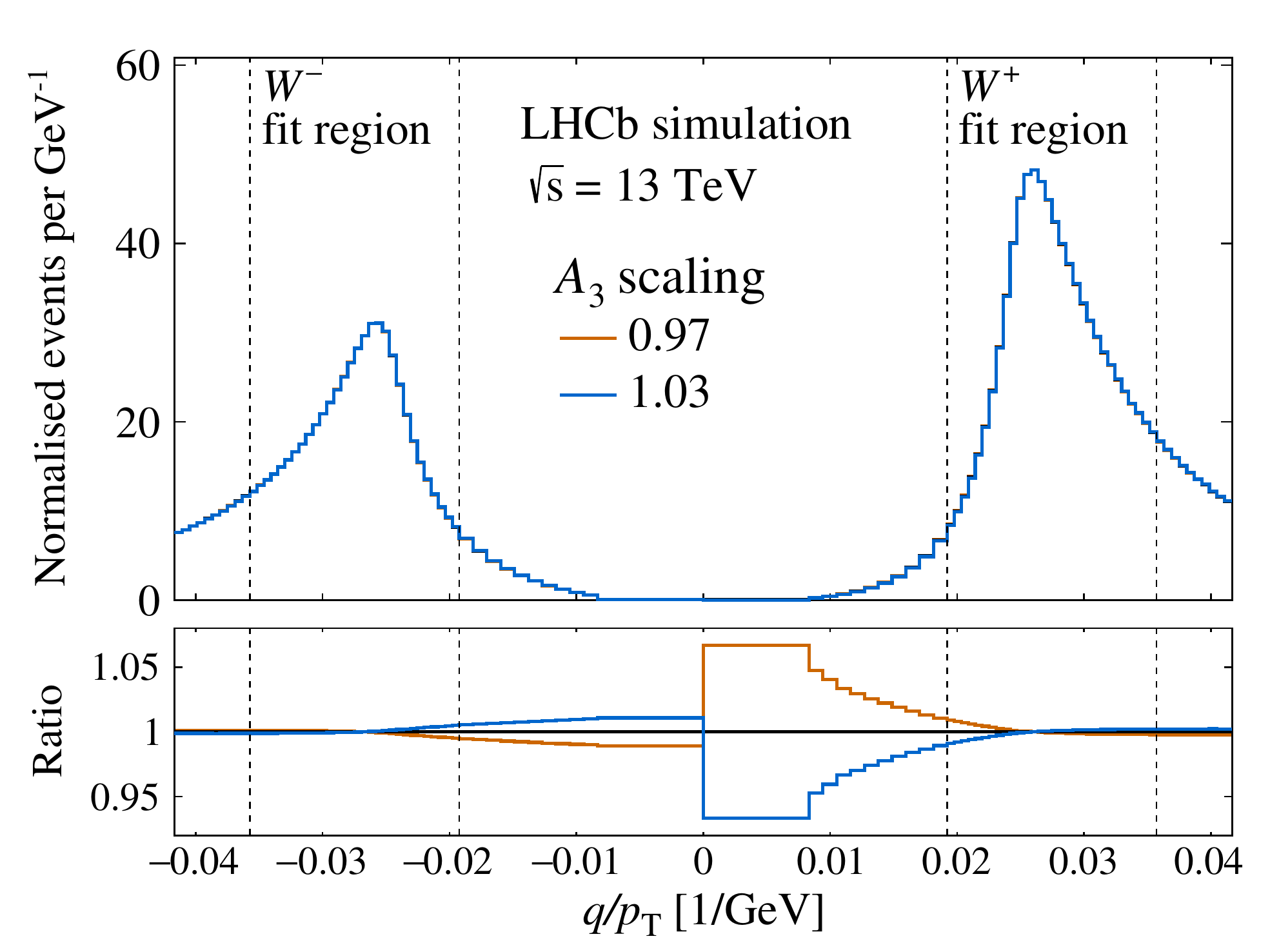}
\caption{\label{fig:Intro}Muon $q/\pt$ distribution in simulated $\PW\to\mu\nu$ events with
variations in (left) $m_W$ and (right) the $A_3$ coefficient. The dashed vertical lines indicate the two \pt ranges that are included in the $m_W$ fit.}
\end{figure}

In this paper a first measurement of $m_W$ is presented using $\PW\to\mu\nu$ decays, including both \PW boson and muon charges, collected at the LHCb experiment.\footnote{The inclusion of charge-conjugate processes is implied throughout unless otherwise specified.} This measurement considers the muon $q/\pt$ distribution, where $q$ is the muon charge.
Figure~\ref{fig:Intro} (left) illustrates how the shape of the muon $q/\pt$ distribution in simulated \PW boson decays is influenced by variations in  $m_W$ of $\pm 300 \mev$,
which corresponds to roughly ten times the target precision of the present analysis.
The $q/\pt$ variable allows all muons with $\pt > 24 \gev$ to be visualised; those with $28 < \pt < 52 \gev$ are used to determine $m_W$, while consistent control of the fit can be demonstrated in the region $\pt > 52 \gev$.

The $\pt$ of a muon produced by the decay of a \PW boson has a strong dependence on the \PW boson transverse momentum ($\pt^W$).
Direct measurements of the $\pt^W$ distribution have been reported by the  ATLAS~\cite{Aad_2012} and CMS~\cite{Khachatryan:2016nbe} collaborations
but the intervals are necessarily coarse due to the limited $\pt^W$ resolution.
Measurements of the transverse momentum distribution for \PZ boson production ($\pt^Z$) are therefore used to validate the predictions for the $\pt^W$ distribution.\footnote{For brevity \PZ denotes $\PZ/\gamma^{\ast}$.}
The angular variable $\phi^{\ast}$~\cite{phistar}, defined in Eq.~\ref{eq:phistar}, is used in this analysis as a proxy for $p_T^Z$ since its distribution can be measured more precisely than that of $p_T^Z$.
Parton-shower programs such as \pythia~\cite{Pythia8Main} can be tuned (\eg Ref.~\cite{MonashTune}) to describe the $\pt^Z$ and $\phi^{\ast}$ data at the per cent level
but it is challenging to reliably translate such tunes to \PW boson production.
However, a \PW-boson-specific tuning of a parton-shower model can be performed simultaneously with a determination of $m_W$~\cite{Lupton}.

If electroweak corrections are neglected then the production and leptonic decay of the \PW boson factorise such that the
differential cross-section can be written as
\begin{align}
  \label{eq:DiffXsec}
\begin{split}
  \frac{\deriv\sigma}{\deriv \pt^W \deriv y \mathrm{d} M \deriv \cos\thetaCS \deriv\phiCS} 
  & = \frac{3}{16\pi}\frac{\deriv\sigma^{\rm unpol.}}{\deriv \pt^W \deriv y \deriv M}\\
  & \big\{(1 + \cos^2\thetaCS) + A_0\frac{1}{2}(1-3\cos^2\thetaCS)  + A_1\sin2\thetaCS\cos\phiCS \\
  & + A_2\frac{1}{2} \sin^2\thetaCS\cos2\phiCS + A_3\sin\thetaCS\cos\phiCS + A_4\cos\thetaCS \\                                                                            
  & + A_5 \sin^2\thetaCS\sin2\phiCS + A_6\sin2\thetaCS\sin\phiCS + A_7\sin\thetaCS\sin\phiCS \big\},\\
\end{split}
\end{align}
where $\thetaCS$ and $\phiCS$ are the lepton decay angles defined in a suitable frame (the Collins-Soper frame~\cite{CollinsSoperFrame1977} is used in this analysis), and  
$\pt^W$, $y$ and $M$ denote the transverse momentum, rapidity and mass of the final state lepton pair, respectively.
An equivalent expression applies to $\PZ \to \mu\mu$ production.
The eight angular coefficients ($A_i$) are ratios of helicity cross-sections and depend on $\pt^W$, $y$ and $M$; $\sigma^{\rm unpol.}$ is usually referred
to as the unpolarised cross-section.
The coefficients $A_5 - A_7$ are numerically small because they only arise at second order, or higher, in the strong coupling constant ($\alpha_s$).\footnote{Throughout this paper $\alpha_s$ denotes the strong coupling at the scale of the \PZ boson mass.}
The coefficient $A_3$ is particularly influential on the muon \pt distribution. Figure~\ref{fig:Intro} (right) shows how the $q/\pt$ distribution in simulated \PW boson events, after the selection requirements described in Sect.~\ref{sec:SelectionCuts}, changes
when $A_3$ is scaled up and down by 3\%.

In this paper, the simulated samples are weighted in the full five-dimensional phase space of vector boson production and decay,
using different models for the unpolarised cross-section, angular coefficients, and QED final-state radiation.
Several PDF sets are used in the analysis but none of the analysed data were included in the determination of these PDF sets.

This paper is organised as follows.
Section~\ref{sec:Samples} describes the data and simulated samples.
Section~\ref{sec:SelectionCuts} details the signal candidate selection requirements.
Section~\ref{section:momentum} describes charge-dependent curvature corrections that are applied to the data and simulation.
The determination of residual smearing corrections to the simulation with a simultaneous fit of $\PZ\to\mu\mu$ and quarkonia decays
is subsequently described.
Section~\ref{sec:Efficiencies} details the measurement of muon selection efficiencies and subsequent weight-based corrections to the simulation.
Section~\ref{sec:qcd-bgd} describes the treatment of background arising from in-flight decays of light hadrons.
Section~\ref{sec:Physics} sets out the modelling of vector boson production and decay.
Section~\ref{sec:MassFit} describes the simultaneous fit of the model to the muon \pt distribution of \PW boson candidates and the $\phi^{\ast}$ (defined in Eq.~\ref{eq:phistar}) distribution of \PZ boson candidates to determine $m_W$.
Section~\ref{sec:Systematics} explains how results based on three different PDF sets are averaged, and summarises the systematic uncertainties.
Several cross-checks of the measurement are reported. The impact of analysis choices and systematic variations on $m_W$ is discussed throughout the paper.
The conclusions of the analysis are presented in Sect.~\ref{sec:Conclusion}. 

\section{Data sets and event selection}
\label{sec:Samples}

The \lhcb detector~\cite{LHCb-DP-2008-001,LHCb-DP-2014-002} is a single-arm forward
spectrometer covering the \mbox{pseudorapidity} range $2<\eta <5$,
designed for the study of particles containing \bquark or \cquark quarks.
The detector includes a high-precision tracking system
consisting of a silicon-strip vertex detector surrounding the proton-proton ($pp$)
interaction region~\cite{LHCb-DP-2014-001}, a large-area silicon-strip detector located
upstream of a dipole magnet with a bending power of about
$4{\mathrm{\,Tm}}$, and three stations of silicon-strip detectors and straw
drift tubes~\cite{LHCb-DP-2017-001} placed downstream of the magnet.
The tracking system provides a measurement of the momentum, \ptot, of charged particles with
a relative uncertainty that varies from 0.5\% at low momentum to 1.0\% at 200\gev.
The minimum distance of a track to a primary $pp$ collision vertex (PV), the impact parameter,
is measured with a resolution of $(15\oplus29/\pt)\mum$,
where \pt is the component of the momentum transverse to the beam, in \gev.
Different types of charged hadrons are distinguished using information
from two ring-imaging Cherenkov detectors~\cite{LHCb-DP-2012-003}.
Photons, electrons and hadrons are identified by a calorimeter system consisting of
scintillating-pad and preshower detectors, an electromagnetic calorimeter and a hadronic calorimeter.
Muons are identified by a system composed of alternating layers of iron and multiwire proportional chambers~\cite{LHCb-DP-2012-002}.
The online event selection is performed by a trigger~\cite{LHCb-DP-2012-004}, 
which consists of a hardware stage, based on information from the calorimeter and muon
systems, followed by a software stage, which applies a full event
reconstruction.

This analysis uses a data sample of $pp$ collisions at $\sqrt{s}=13$~TeV recorded during 2016, corresponding to an integrated luminosity of about $1.7\invfb$.
Roughly half of the data were recorded in each of the dipole magnet polarity configurations,
resulting in a large degree of cancellation of charge-dependent curvature biases and their associated uncertainties.
These data correspond to an average number of proton-proton interactions per bunch-crossing event of $\mathcal{O}(1)$.

During Run~2 the LHCb detector was aligned and calibrated in real-time~\cite{LHCb-DP-2019-001}.
The alignment of the tracking system is based on a $\chi^2$ minimisation of the residuals of the clusters of tracker hits evaluated with a Kalman filter that
takes into account multiple scattering and energy loss~\cite{Hulsbergen_2009}.
The alignment algorithm also permits mass and vertex constraints~\cite{Amoraal_2013}.
An optimised offline alignment, which includes \Zmm events with a mass constraint that accounts for the natural width of the \PZ boson, is used to determine the track parameters.
Since the real-time alignment is not optimised for the analysis of high-\pt final states, this realignment improves the \Zmm mass resolution by around 30\%.

All signal and background processes are simulated using an LHCb specific tune~\cite{LHCb-PROC-2010-056} of \pythia version 8.186~\cite{Pythia8Main}.
  The interaction of the generated particles with the detector, and its response,
  are implemented using the \geant
  toolkit~\cite{Allison:2006ve, *Agostinelli:2002hh} as described in
  Ref.~\cite{LHCb-PROC-2011-006}.
Events are simulated with both polarity configurations and weights are assigned to
events in each polarity such that the polarity distribution matches the recorded data.

\section{Selection of \boldmath{\PW} boson, \boldmath{\PZ} boson and quarkonia signal candidates}
\label{sec:SelectionCuts}

Tracks are identified as muons if they are matched to hits in either three or all four of the most downstream muon stations depending on their momentum. They are then considered in this analysis if they are within the range $1.7 < \eta < 5.0$,
and have a momentum of less than $2\tev$.
The tracks must have a good fit quality and a relative momentum uncertainty of less than 6\%.

Candidate $W\to\mu\nu$ events are selected online by requiring that one identified muon
satisfies the requirements of all stages of the trigger.
At the hardware stage a \pt of at least 6\gev is required.
The isolation of a muon is defined as the scalar sum of the transverse momenta
of all charged and neutral particles, as selected by a particle-flow algorithm, described in Ref.~\cite{LHCb-PAPER-2013-058}, within $(\Delta \eta)^2 + (\Delta \phi)^2 < 0.4^2$ around the muon, where $\Delta\eta$ and $\Delta\phi$ denote the separation in $\eta$ and azimuthal angle around the beam direction ($\phi$), respectively.
Hadronic background contributions are suppressed by requiring the muon to have an isolation of less than 4\gev.
For the $\PW\to\mu\nu$ selection the $\eta$ range is tightened to $2.2 < \eta < 4.4$ so that the area of the isolation cone is fully instrumented.
The muon must satisfy $\chisqip < 9$ where \chisqip is defined by the difference in the vertex fit $\chi^2$ of the PV with and without including the muon.
Background from \PZ boson events is suppressed by rejecting events that contain a second muon with $\pt > 25\gev$ and an opposite charge to that of the primary muon candidate. Roughly $2.4$~million $\PW\to\mu\nu$ candidates are selected in the range $28 < \pt < 52$~GeV.

Candidate $\PZ\to\mu\mu$ events are reconstructed
from combinations of two oppositely charged identified muons associated to the same PV
with an invariant mass within $\pm 14$\gev of the known \PZ boson mass~\cite{PDG2020}.
At least one muon must be matched to a single muon selection at all stages of the trigger.
Both muons must have $\pt > 20$\gev, an isolation value below 10\gev, 
and an impact parameter significance of less than ten standard deviations.
Roughly $190$~thousand $\PZ\to\mu\mu$ candidates are selected.

Candidate $\jpsi\to\mu\mu$ and $\Upsilonres(1S) \to\mu\mu$ events, which are primarily used to calibrate the modelling of the momentum measurement, are required to have a pair of oppositely charged identified muons.
Both muons must have a transverse momentum above 3\gev and satisfy a tighter muon identification requirement.
In order to specifically select $\jpsi\to\mu\mu$ candidates originating from $b$-hadron decays the decay vertices must be displaced from the nearest PV with a significance of at least three standard deviations. These selections retain roughly $1.0$ million $\Upsilonres(1S) \to\mu\mu$ candidates and $220$ thousand $\jpsi\to\mu\mu$ candidates.

\section{Momentum calibration and modelling} 
\label{section:momentum}

\begin{figure}[!t!]
  \centering
  \includegraphics[width=0.49\textwidth]{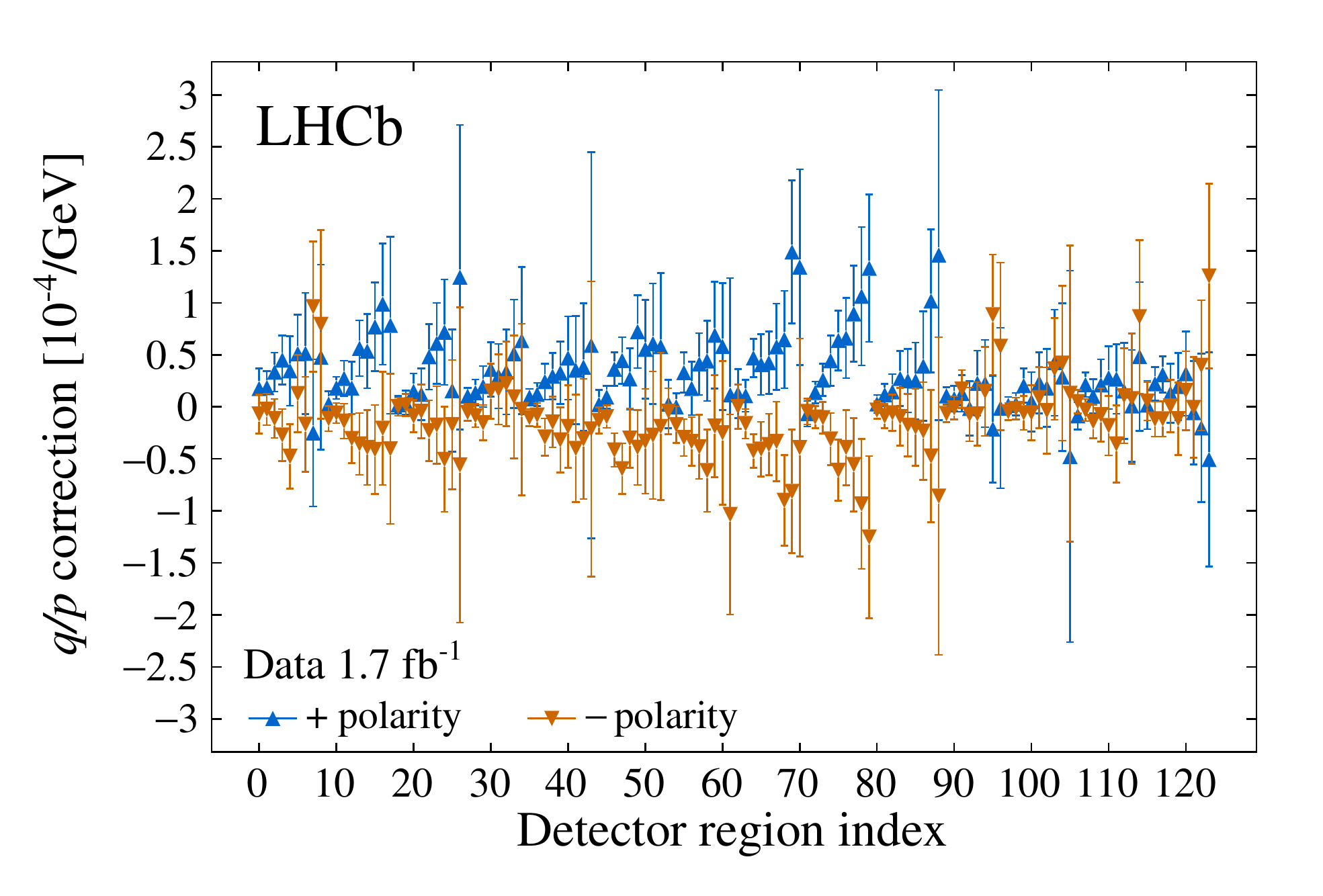}
  \includegraphics[width=0.49\textwidth]{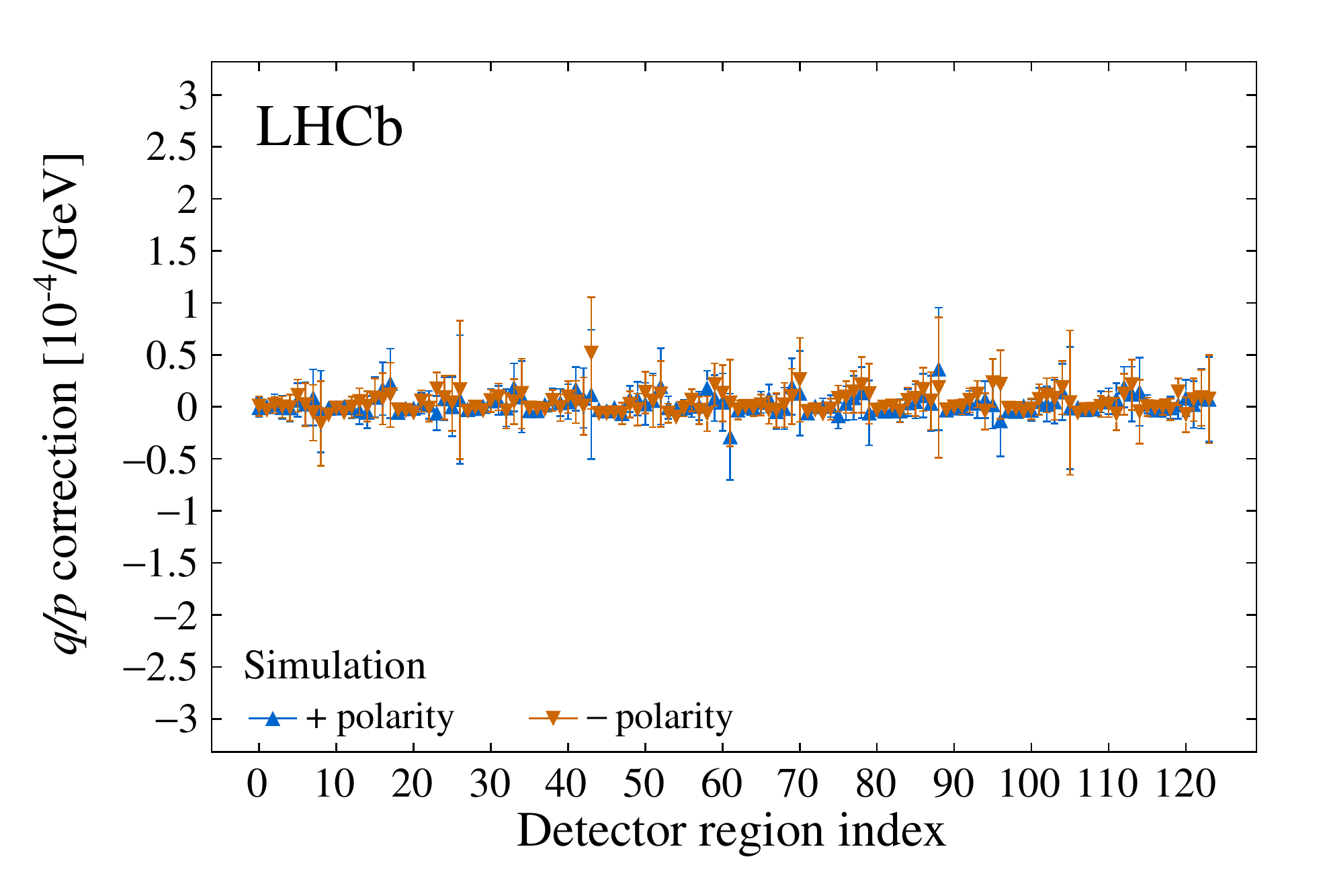}
 \caption{Curvature corrections as a function of the detector region index (depends on $\eta$, $\phi$ and tracking detector, as described in the text) for (left) data and (right) simulation. The corrections are shown for both polarity configurations. The periodic pattern corresponds to a dependence on pseudorapidity that repeats in the intervals of the azimuthal angle.}
  \label{fig:alignment_maps}
\end{figure}

The momentum scale can be precisely determined from the mass measurements of various resonances, including those that decay to muon pairs.
However, charge-dependent curvature biases that shift $q/p$ are challenging to
estimate because their effect largely cancels in the mass of the resonances.
They are also particularly important for the high momentum muons from \PW and \PZ boson decays.
In Ref.~\cite{Barter:2021npd} it was proposed to determine corrections using the so-called \emph{pseudomass} variable in \Zmm events
\begin{equation}
\mathcal{M}^\pm = \sqrt{2p^\pm \pt^\pm \frac{p^\mp}{\pt^\mp}(1-\cos\theta)},\label{eq:pseudomassdef}
\end{equation}
where $p^{\pm}$ and $\pt^{\pm}$ are the momenta and transverse momenta of the $\mu^{\pm}$, respectively.
The opening angle between the two muons is denoted $\theta$.
Crucially, the value of $\mathcal{M^{\pm}}$ is independent of the magnitude of the momentum of the $\mu^{\mp}$ and is therefore
directly sensitive to curvature biases affecting the $\mu^{\pm}$ candidate.
The pseudomass is an approximation of the dimuon mass under the assumption that the dimuon system has
zero momentum transverse to the bisector of the two lepton transverse momenta.
The $\phi^{\ast}$ observable is defined as~\cite{phistar}
\begin{equation}\label{eq:phistar}
\phi^{\ast} = \frac{\tan((\pi-\Delta\phi)/2)}{\cosh(\Delta\eta/2)} \sim \frac{\pt^Z}{M},
\end{equation}
where $\Delta\phi$ is the azimuthal opening angle
between the two leptons and $\Delta\eta$ is the difference between the pseudorapidities of the
negatively and positively charged lepton.
In events with small values of $\phi^{\ast}$ the pseudomass better approximates the dimuon mass.
The pseudomass distributions for events with $\phi^{\ast} < 0.05$ are studied in intervals of $\phi$ and $\eta$ of the $\mu^{\pm}$ candidate, with a further categorisation into candidates traversing the silicon strip or straw drift tube detectors downstream of the magnet.
A maximum likelihood fit of the $\mathcal{M^{\pm}}$ distributions is performed for each of these detector regions.
The signal shapes are described by the sum of a resonant Crystal-Ball~\cite{Skwarnicki:1986xj} component
and a nonresonant component represented by an exponential function.
The means of the $\mathcal{M^{\pm}}$ Crystal-Ball functions are parameterised as $\bar{M}(1 \pm \mathcal{A})$, where $\mathcal{A}$ and $\bar{M}$ are freely varying asymmetry and mass parameters, respectively. %related by a freely varying asymmetry parameter $\mathcal{A}$.
The resulting $q/p$ corrections, which are given by 
$\mathcal{A}/\bar{p}$ where $\bar{p}$ is the average muon momentum for a given interval in $\eta$ and $\phi$, are presented for the data and simulation for both polarity configurations in Fig.~\ref{fig:alignment_maps}.

\def\sigmaA     {{\ensuremath{\sigma_\text{MS}}}\xspace}
\def\sigmaB     {{\ensuremath{\sigma_{\delta}}}\xspace}

After the curvature corrections are applied to the data and simulation,
the momenta of the simulated muons are smeared to match those in the data, as described below, according to
\begin{equation}\label{eq:smearing}
\frac{q}{p} \to \frac{q}{p\cdot\mathcal{N}(1+\alpha,\sigmaA)} + \mathcal{N}\left(\delta ,\frac{\sigmaB}{\cosh\eta}\right),
\end{equation}
where $\mathcal{N}(a,b)$ represents a random number sampled from a Gaussian distribution with mean $a$ and width $b$.
The \sigmaA and \sigmaB parameters correspond to the multiple scattering and curvature measurement contributions to the resolution, respectively.
The smearing model includes six parameters in total.
There are two momentum scale parameters $\alpha$ corresponding to the $2.2 < \eta < 4.4$ region, which coincides with the selection of \PW boson candidates, and the $\eta < 2.2$ region.
A single $\delta$ parameter, corresponding to a curvature bias, covers the region $2.2 < \eta < 4.4$,
while the value of $\delta$ is fixed to zero in the region $\eta < 2.2$.
There are two $\sigmaB$ parameters corresponding to the $2.2 < \eta < 4.4$ and $\eta < 2.2$ regions,
while a single $\sigmaA$ parameter is found to adequately cover all $\eta$ values.
The empirical $1/\!\cosh\eta$ dependence of the second term in Eq.~\ref{eq:smearing} improves the modelling of the $\eta$ dependence in the \Zmm mass distribution. 
As a further correction to Eq.~\ref{eq:smearing}, the value of \sigmaA is increased by a factor of 1.5 in the region $\eta > 3.3$ since this improves the agreement between data and simulation in the $\eta$ dependence of the quarkonia mass distributions.

The six smearing parameters are determined in a  simultaneous fit of $\jpsi\to\mu\mu$, $\Upsilonres(1S)\to\mu\mu$ and $Z\to\mu\mu$ candidates in data and simulation.
A total of 36 dimuon invariant mass distributions are used in the fit.
First, there are three $\eta$ regions covering $\eta < 2.2$, $2.2 < \eta < 3.3$ and $3.3 < \eta < 4.4$,
which result in six categories that depend on the $\eta$ regions of the two muons.
The quarkonia mass distributions are only used in categories with both muons having $\eta > 2.2$.
In the subset of the $\eta$ categories with both muons in $\eta > 2.2$, the \Zmm data are split into three intervals of the asymmetry between the momenta of the two muons,
which provides a first order sensitivity to the $\delta$ parameters.
Finally, all categories are divided by magnet polarity.

As in previous studies of \Zmm production with the LHCb experiment~\cite{LHCb-PAPER-2016-021}, the background under the \Zmm peak is low enough to be neglected but the fit includes exponential
functions for the background contributions under the quarkonia resonance peaks.
The fractions and slopes of these exponential components vary freely in the fit.

\begin{figure}[!b!]\centering
  \includegraphics[width=\DefaultFigWidth]{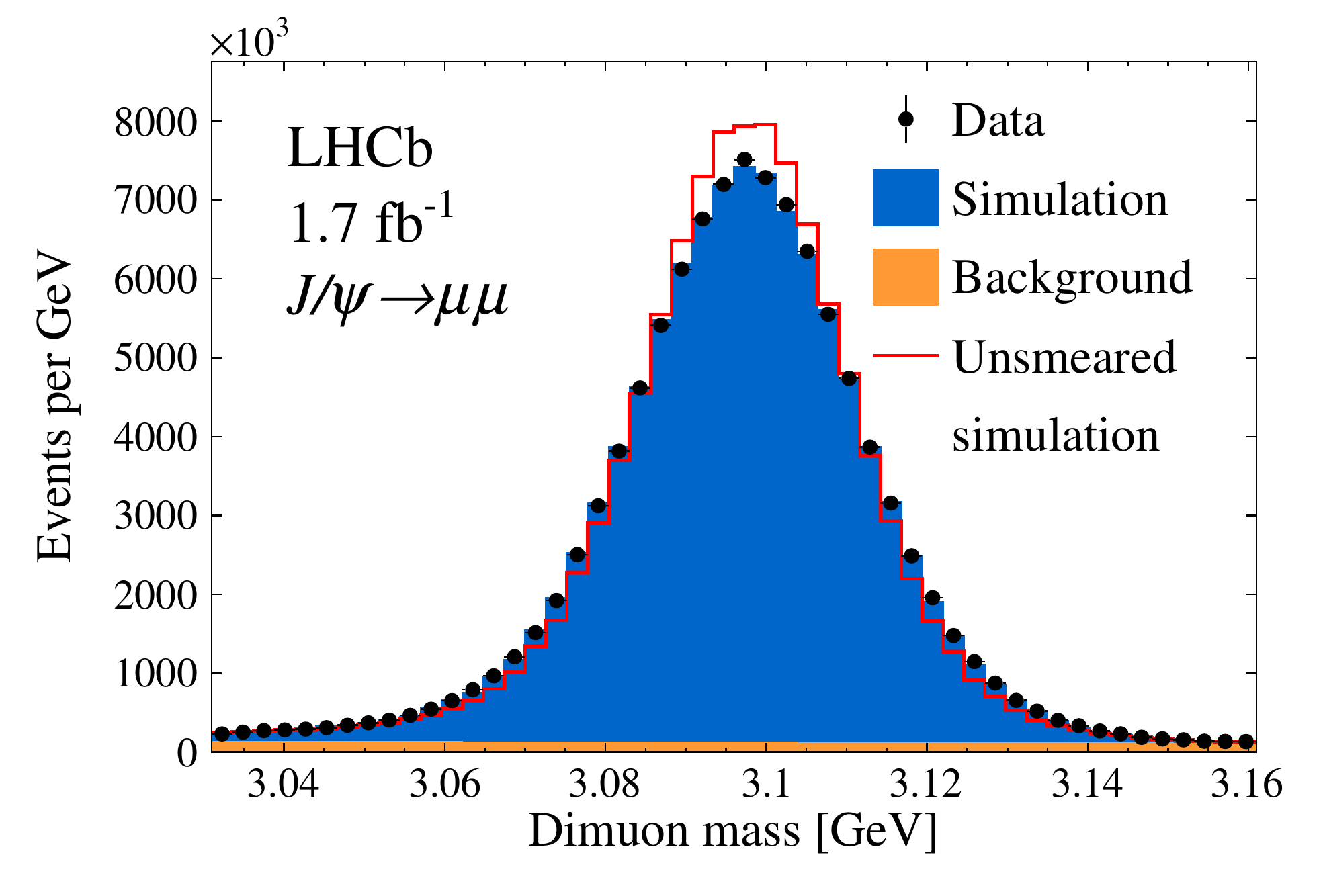}
  \includegraphics[width=\DefaultFigWidth]{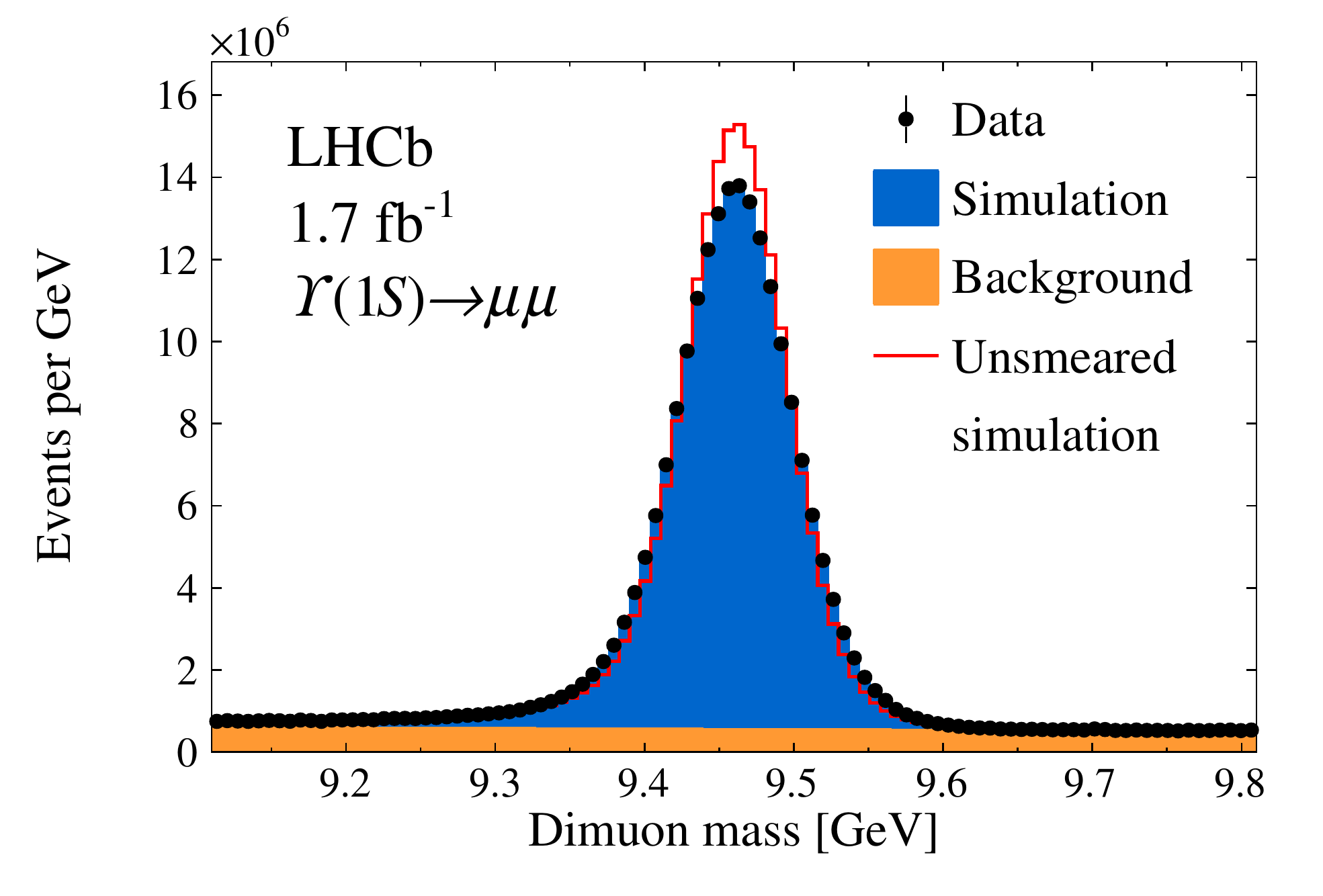}
  \includegraphics[width=\DefaultFigWidth]{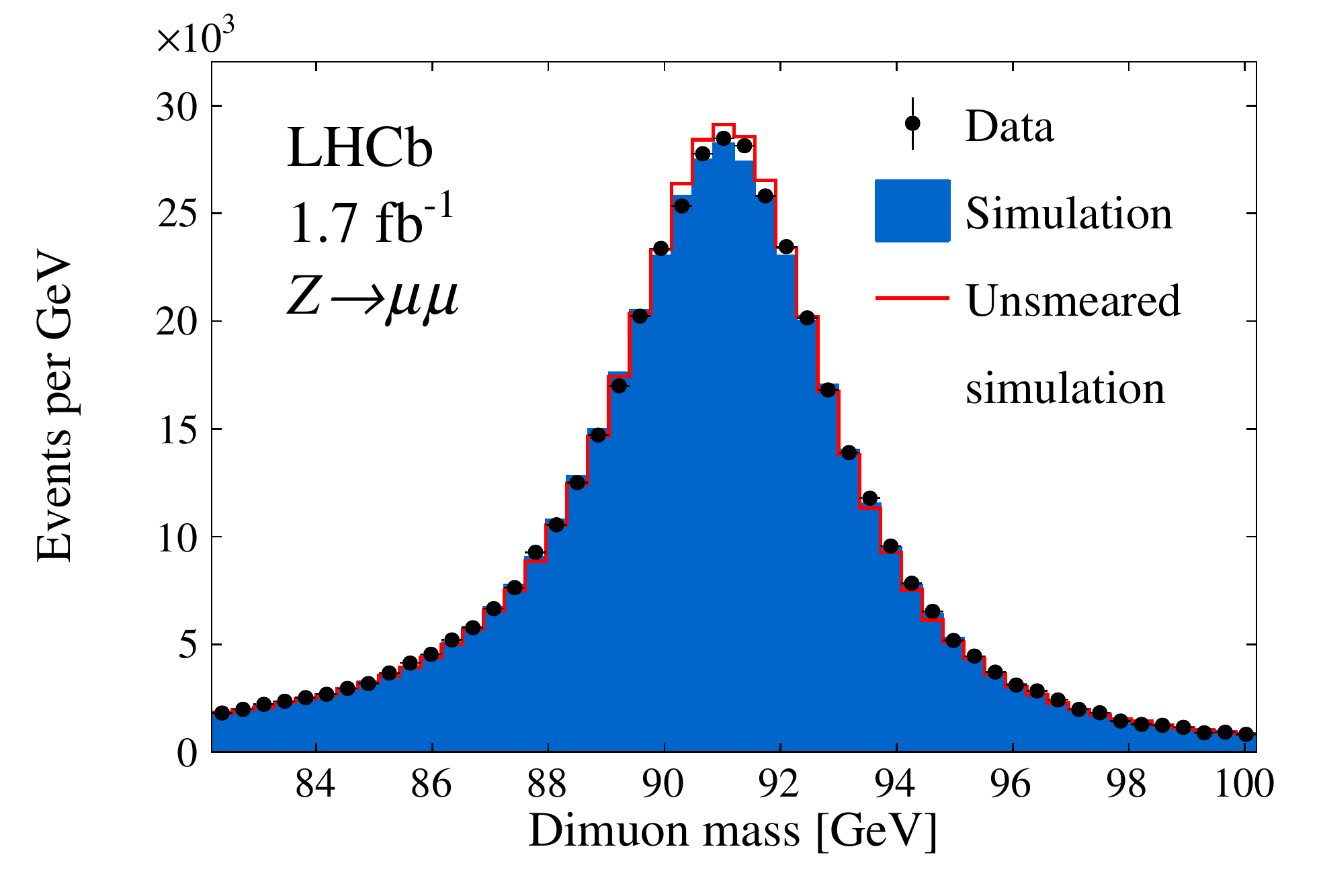}
  \caption{Dimuon mass distributions for selected $J/\psi$, $\Upsilonres(1S)$ and \PZ boson candidates.
    All categories with both muons in the $2.2 < \eta < 4.4$ region are combined.
    The data are compared with the fit model.
    The red histogram delineates the model before the application of the smearing.}
\label{fig:MomentumScaleFit}
\end{figure}

The total $\chi^2$ from the fit is 1862 for 2082 degrees of freedom.
Table~\ref{tab:MomentumScale} shows the fit values of the six parameters in the smearing model.
The $\delta$ values are close to zero as expected given the curvature corrections that have already been applied.
Figure~\ref{fig:MomentumScaleFit} shows the dimuon mass distributions for the \PZ, $\Upsilonres(1S)$ and \jpsi samples after combining all categories with
both muons in the $2.2 < \eta < 4.4$ range.

\begin{table}[]\centering
  \caption{\label{tab:MomentumScale}Parameters in the momentum smearing model where the uncertainties quoted are statistical.}
  \begin{tabular}{ll}
    Parameter & Fit value \\
    \hline
    $\alpha$ ($\eta < 2.2$) & $(0.58 \pm 0.10)\times 10^{-3}$\\
    $\alpha$ ($2.2 < \eta < 4.4$) & $(-0.0054 \pm 0.0025)\times 10^{-3}$\\
    $\delta$ & $(-0.48 \pm 0.37)\times 10^{-6}~\gev^{-1}$\\
    $\sigmaB$ ($\eta < 2.2$) & $(17.7 \pm 1.2)\kev^{-1}$\\
    $\sigmaB$ ($2.2 < \eta < 4.4$) & $(14.9 \pm 0.9)\kev^{-1}$\\
    $\sigmaA$ & $(2.015 \pm 0.019)\times 10^{-3}$\\
    \hline
  \end{tabular}
\end{table}

The statistical uncertainties in the smearing parameters result in an uncertainty in $m_W$ of $3\mev$.
The uncertainty in the world average of the $\OneS$ mass~\cite{PDG2020} leads to an uncertainty of $2\mev$.
The uncertainty in the $\jpsi$ mass is negligible compared to that in the $\OneS$ mass. The $\OneS$ and the \PZ masses have comparable relative uncertainties but the latter has a negligible effect given the limited size of the \PZ boson sample.
The amount of material in the detector, which affects the modelling of energy losses, is varied by 10\%, leading to
an uncertainty of $3\mev$.
A total uncertainty of $5\mev$ is attributed to the shifts in $m_W$ corresponding to: an alternative form of the $1/\!\cosh\eta$ factor in Eq.~\ref{eq:smearing}; and a variation in the $\eta$ region over which \sigmaA is scaled.
An uncertainty of $2\mev$ is attributed to the modelling of the radiative tails in the \OneS and \jpsi simulation, using the methods described in Ref.~\cite{LHCb-PAPER-2013-016}.
The total uncertainty attributed to the modelling of the momentum scale and resolution is $7\mev$.

\section{Efficiency corrections}
\label{sec:Efficiencies}

Corrections to the simulation are required for the muon trigger, identification, tracking and  isolation efficiencies.
The efficiencies are measured using a combination of \Zmm and, in the case of the trigger efficiency, \Upsmumu samples.
Positively and negatively charged muons are analysed separately but the results are combined since any charge asymmetries
are verified to have a negligible effect.

The trigger efficiency, which accounts for the hardware and software stages, is measured using a combination of \Zmm and \Upsmumu events
in which one so-called {\em tag} muon is required to match a positive decision in the hardware trigger and the first stage of the software-level trigger such that the other muon
can be regarded as an unbiased {\em probe} of the trigger efficiency.
Events are categorised as either {\em matched} or {\em unmatched} depending on whether the probe muon is matched to a positive trigger decision in the event data record.
The \Zmm sample is verified to be sufficiently pure that the efficiencies can be measured by simply counting the matched and unmatched events with invariant masses within $\pm 15$\gev of the known \PZ boson mass~\cite{PDG2020}.
The efficiencies are determined in four uniform $\phi$ intervals and eight uniform $\eta$ intervals in the range $2.2 < \eta < 4.4$.
There are two additional $\eta$ intervals in the region $\eta < 2.2$ and one in the region $\eta > 4.4$.
The \Upsmumu sample requires background subtraction by fitting the dimuon invariant mass distribution with a parametric model of the signal and background components.

\begin{figure}[!b!]
  \centering
  \includegraphics[width=0.9\textwidth]{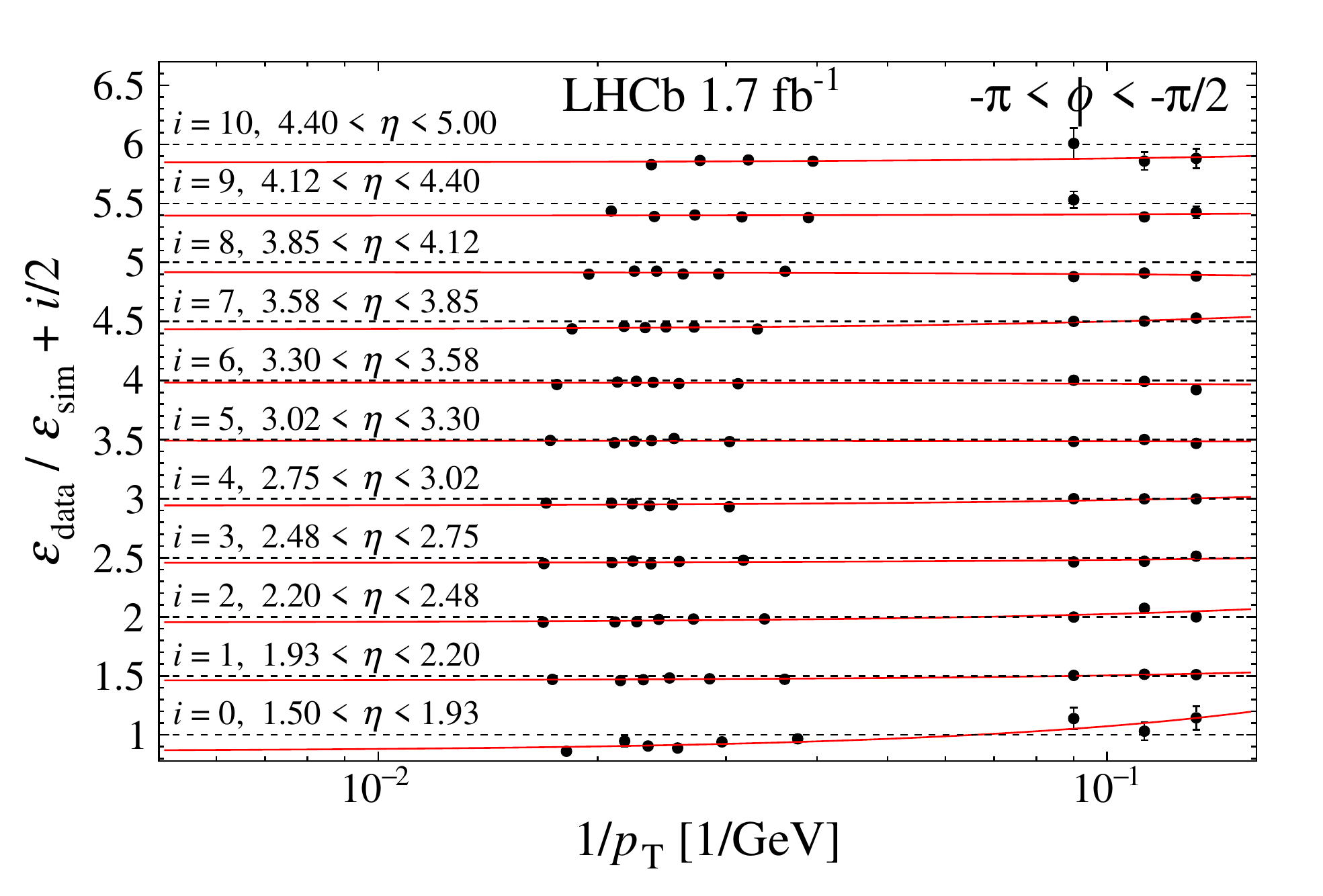}
  \caption{Trigger efficiency ratios in data relative to simulation in intervals of $\eta$ for a single interval of $\phi$. The points are presented as a function of $1/\pt$ and are overlaid with best-fit linear functions of $\pt$. Points from each $\eta$ interval are separated for readability by an offset of half the integer index $i$ for that interval. The uncertainties on most of the points are too small for the error bars to be visible.}    
  \label{fig:L0HLTEfficiencyRatios}
\end{figure}

Three $\pt$ intervals, in the range $7.0 < \pt < 12.5$\gev, are used for the probe muons from \Upsmumu decays while for the $\PZ\to\mu\mu$ candidates an adaptive algorithm is employed to determine the \pt intervals.
The ratios of the trigger efficiencies in data relative to those in the 
simulation are shown as a function of $1/\pt$ of the muon in Fig.~\ref{fig:L0HLTEfficiencyRatios} for each of the intervals in $\eta$ and $\phi$.
These are overlaid with a linear function of \pt, from which correction weights for the simulated events are evaluated.
The weights for  the \PW boson model only rely on these functions but the weights for the \PZ boson model also require
a parameterisation of the absolute efficiency in the simulation such that the efficiency can be correctly modelled for \PZ boson candidates with one or two muons matched to a trigger decision. The absolute efficiency is described by an error function that captures the \pt threshold (roughly 6\gev) of the hardware trigger~\cite{LHCb-DP-2019-001}.

The muon identification efficiency is treated in a similar manner to the trigger efficiencies, using \Zmm events. 
The resulting event weights, which are applied to the simulated events, are within a few per cent from unity.
The tracking efficiency is determined as in previous measurements of \PW and \PZ boson production at LHCb~\cite{LHCb-PAPER-2016-021} using \Zmm candidates where the probe muons are reconstructed by combining hits from the muon stations and the large-area silicon-strip detector located upstream of the magnet~\cite{LHCb-DP-2013-002}. As neither of these detectors are used in the primary track reconstruction algorithms, the probes can be used to measure the tracking efficiency. Correction factors are evaluated using a similar approach to those of the muon identification efficiency,
except that the corrections are assumed to be independent of $\pt$.

The statistical uncertainties in the muon trigger, tracking and identification efficiency corrections are evaluated by
rerunning the relevant steps of the analysis, up to and including the $m_W$ determination, with random fluctuations in the underlying efficiency values.
The RMS of the resulting variations in the $m_W$ value is regarded as an uncertainty.
A systematic uncertainty is attributed to the dependence of the results on the scheme for $\eta$ and $\phi$ intervals.
This includes restricting to a single interval in $\phi$, reducing the number of $\eta$ intervals (within $2.2 < \eta < 4.4$) by a factor of two,
varying the number of $\pt$ intervals between two and ten, and using the simulation rather than the data to control the adaptive algorithm.
Further systematic uncertainties are attributed to variations in the isolation and \pt requirements on the tag muons,
the mass windows used to determine the \Zmm signal yields, and the functions of the \pt dependence of the efficiency ratios.
As the probe muons for the tracking efficiency are reconstructed using minimal tracking information, they have a significantly lower momentum resolution and so a dedicated momentum smearing is applied by default. A variation in the
size of this smearing is included in the systematic uncertainty evaluation.
The total uncertainty associated to the muon trigger, identification and tracking uncertainties
is $6\mev$.

\begin{figure}[!b!]\centering
  \includegraphics[width=\DefaultFigWidth]{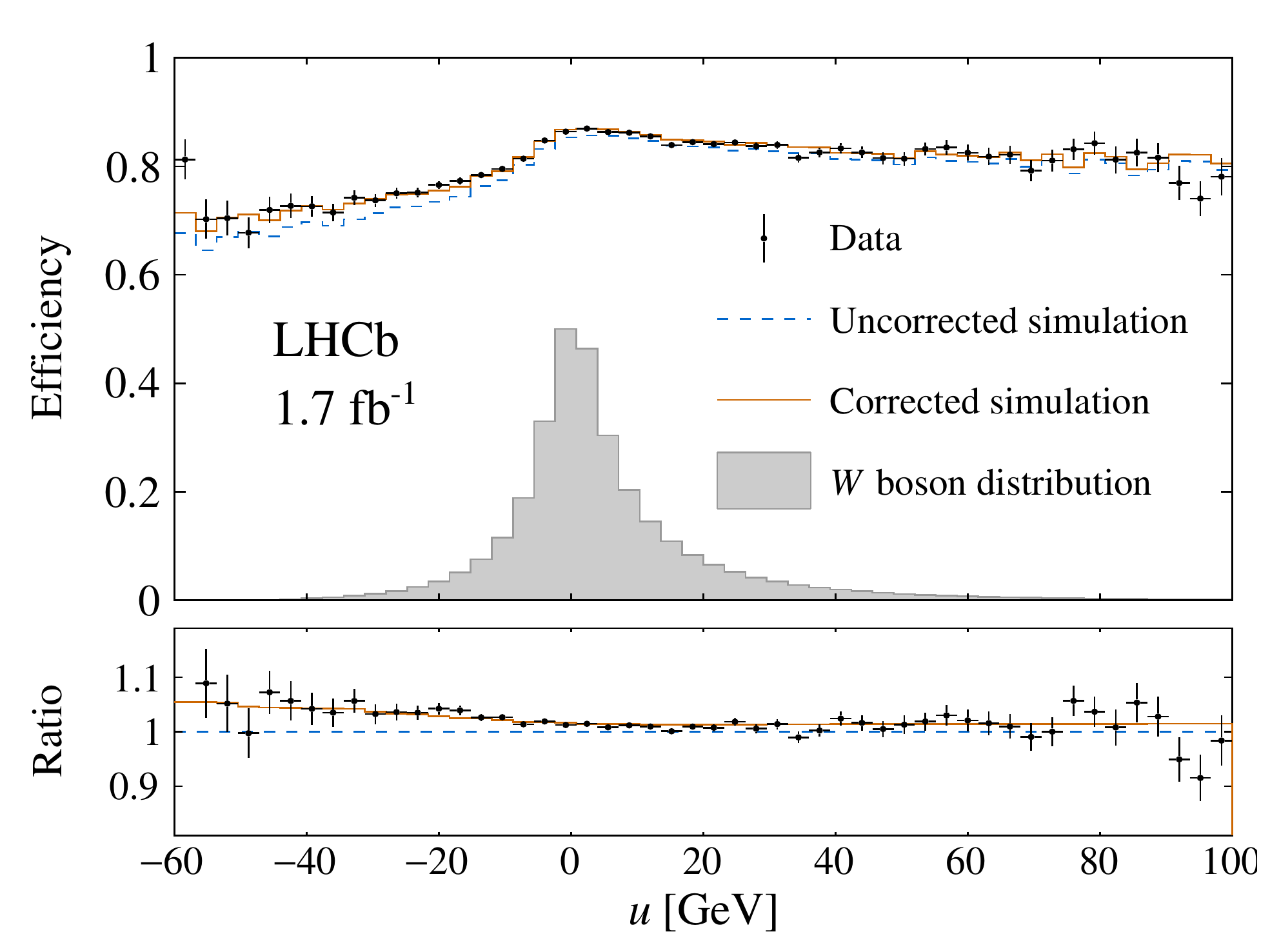}
  \includegraphics[width=\DefaultFigWidth]{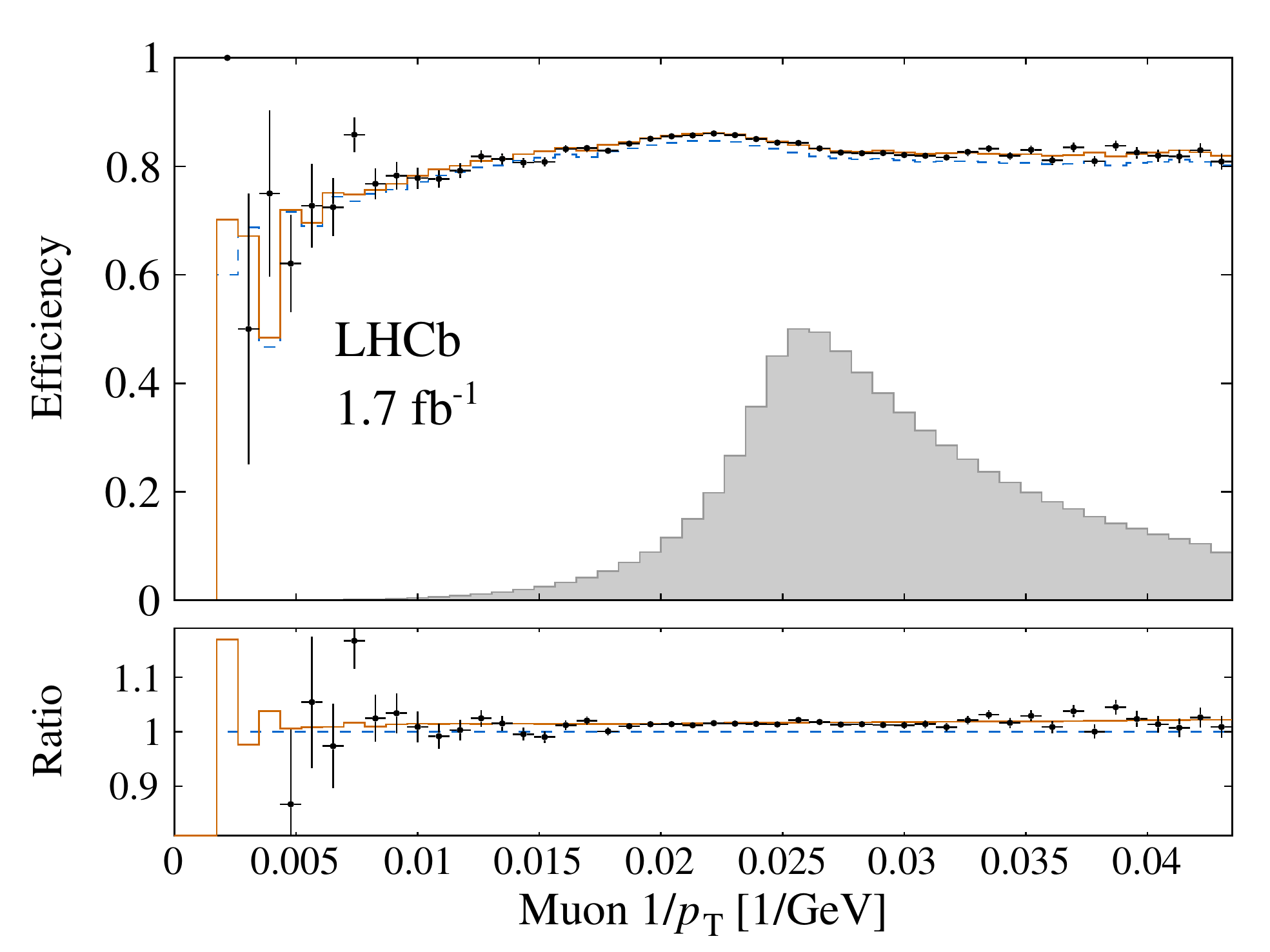}
  \caption{Isolation efficiencies as a function of the observables (left) $u$ and (right) $1/\pt$ of the muon. The grey histograms indicate the, arbitrarily normalised, shapes of each distribution in simulated \PW boson events. 
  In the lower panels the ratios of the isolation efficiency with respect to the uncorrected simulation are shown.} \label{fig:IsolationValidation1}
\end{figure}

The efficiency of the isolation requirement is measured with \Zmm events.
The isolation variable receives contributions from pile-up, the underlying event and the recoil component of the hard process.
The {\em recoil projection} for each muon is defined by 
\begin{equation}
u = \frac{\vec{p}^{\;V}_{\rm T} \cdot \vec{p}^{\;\mu}_{\rm T}} {p_{\rm T}^{\mu}},
\end{equation}
where $\vec{p}^{\;\mu}_{\rm T}$ and $\vec{p}^{\;V}_{\rm T}$ are the two-dimensional momentum vectors of the muon and the parent vector boson in the transverse plane.
Figure~\ref{fig:IsolationValidation1} (left) shows the isolation requirement efficiency as a function of $u$ in the \Zmm data and simulation.
The efficiency is around 80\% at positive values of $u$ where the underlying event contribution dominates.
At negative values of $u$, corresponding to large recoil, the efficiency drops to around 70\%.
The full reconstruction of the \PZ boson in \Zmm events allows the determination of corrections as a function of $u$ defined at reconstruction level and consistently applied to \PW and \PZ boson events as a function of $u$ defined at generator level, with this approach validated using the \PZ boson sample.
A map of relative efficiencies between data and simulation is determined in intervals of $u$ and $\eta$
and is used to evaluate weights for the simulated events.
Figure~\ref{fig:IsolationValidation1} (right) shows that the dependence of the isolation efficiency on $1/\pt$ in data is accurately described by the simulation after the corrections.
The statistical uncertainties in the isolation efficiency corrections are treated as a source of systematic uncertainty.
This is combined in quadrature with a systematic uncertainty that accounts for variations in the $u$ and $\eta$ intervals
and in a smoothing procedure applied to enhance the effective statistical precision of the correction map.
The total uncertainty attributed to the isolation efficiency modelling is $4\mev$.

The modelling of the impact parameter and track fit $\chi^2$ variables in simulation is improved in two stages, both of which make use of \Zmm events.
Initially, the values of the variables in the simulation are smeared and shifted to match the data.
Subsequently, weights are applied to the simulated events to correct for small residual differences in the 
efficiencies of the selection requirements between data and simulation.
In order for the impact parameter modelling to be reliably transported between \PZ and \PW boson events,
the PVs are refitted with all signal muons removed.
The three-dimensional impact parameter is then decomposed into its individual components and these are smeared according to a normal distribution in six intervals in $\eta$ and seven intervals in $\phi$.
A similar procedure is used to improve the modelling of the track fit $\chi^2$ distribution.
These corrections are followed by smaller corrections applied to account for the efficiencies of the impact parameter and track $\chi^2$ requirements.
The efficiency weights are also determined with \Zmm events and are typically within a few per mille from unity.
Neither the statistical uncertainties nor reasonable variations in the $\eta$ and $\phi$ interval schemes are found to have significant impact on the $m_W$ value. Therefore, no systematic uncertainty is considered.

\section{QCD background model}\label{sec:qcd-bgd}

\begin{figure}[tbp]\centering
  \includegraphics[width=\DefaultFigWidth]{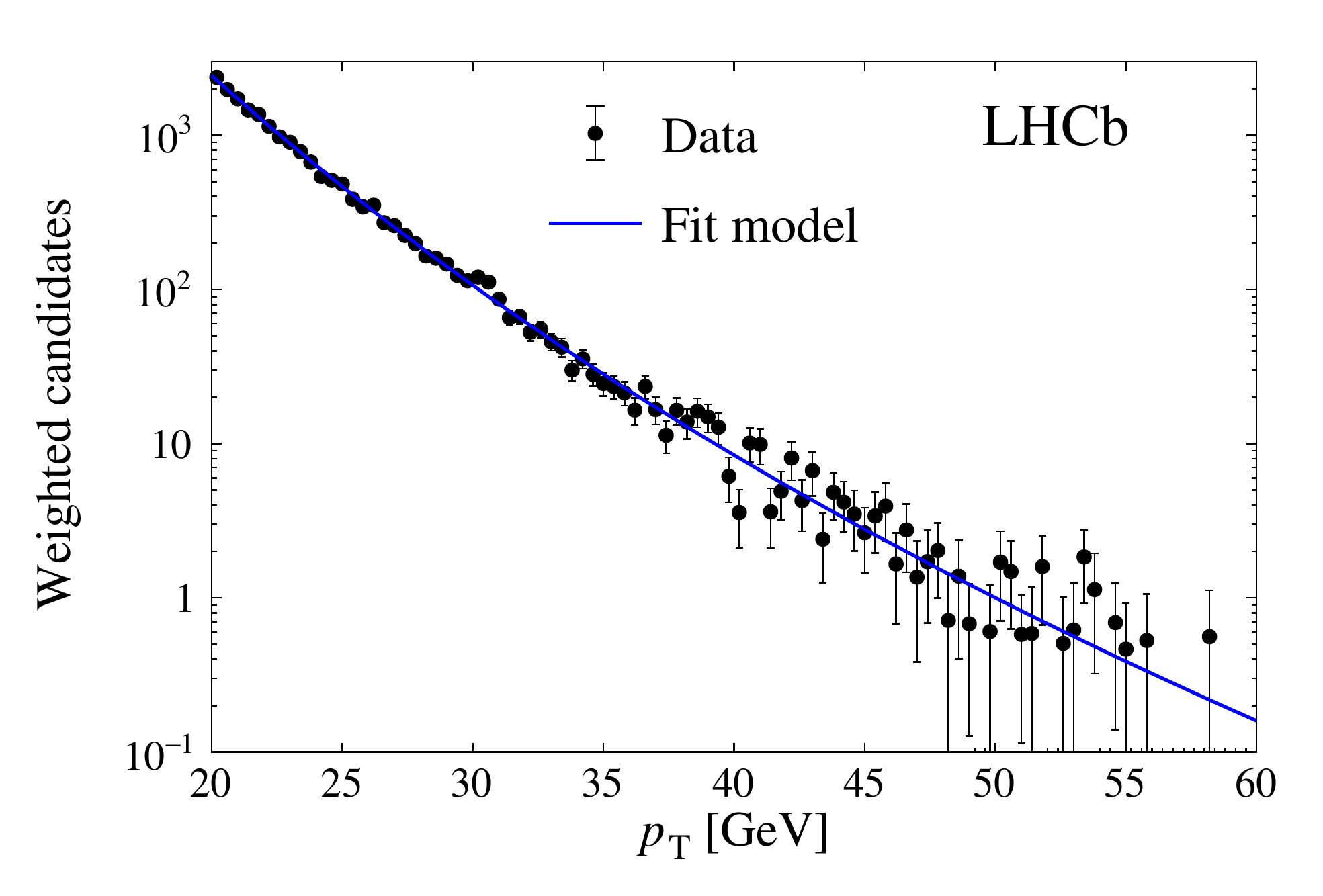}
  \includegraphics[width=\DefaultFigWidth]{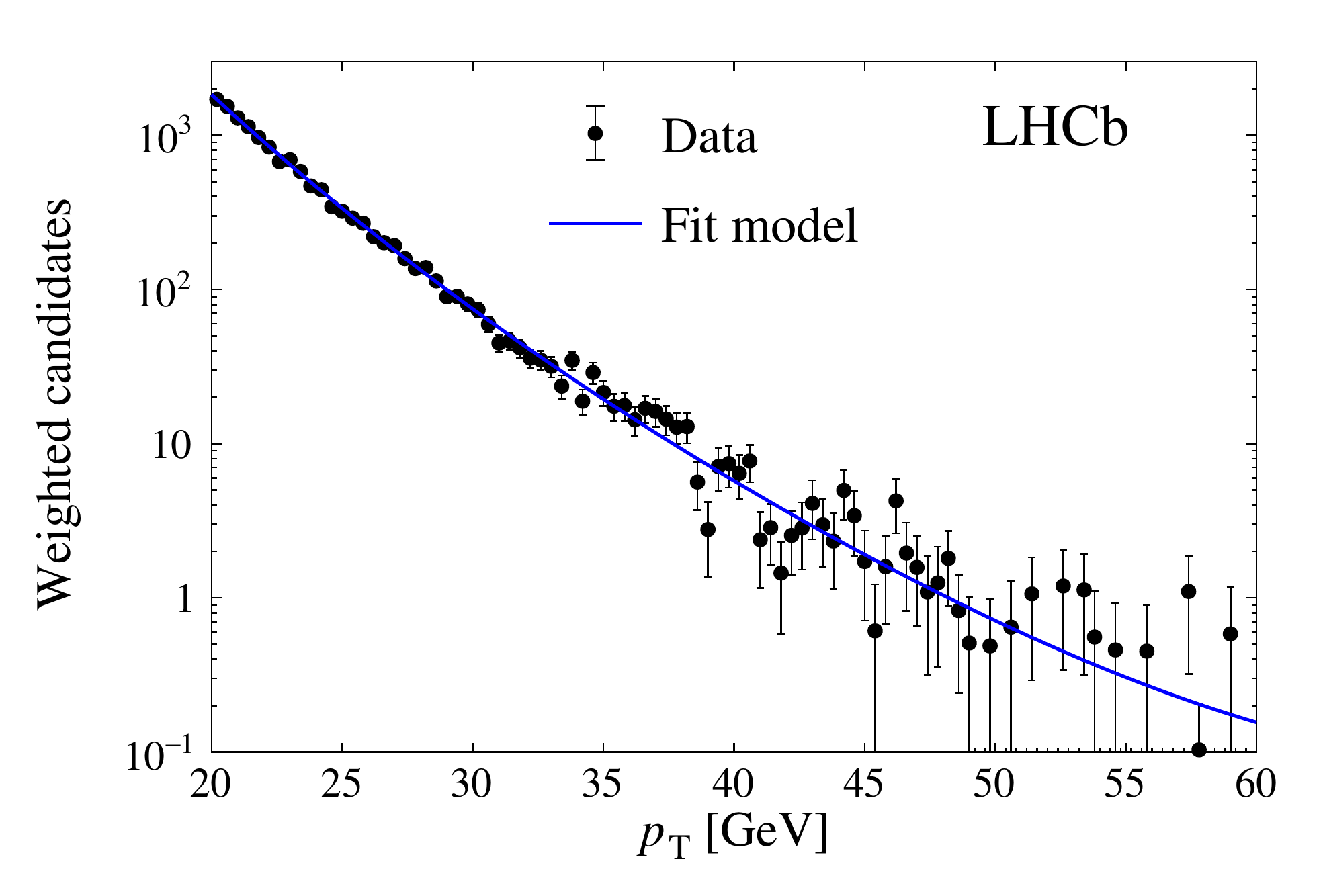}
  \caption{Weighted \pt spectra of the samples of (left) positively and (right) negatively charged hadron candidates. The fit results are overlaid.}
  \label{fig:qcd_bgd_fit}
\end{figure}

A small background from in-flight decays of pions and kaons into muons is present in the sample of $\PW \to\mu\nu$ boson candidates.
This background cannot be modelled with high enough accuracy using full detector simulation.
It is therefore modelled using a sample of high-$\pt$ tracks, selected by dedicated triggers without muon identification requirements.
The \PW boson selection requirements are applied to this sample but with the muon identification requirement inverted.
The resulting sample is verified in simulation to be a pure sample of charged hadrons, composed of roughly 60\% pions, 30\% kaons and 10\% protons, produced directly at the  $pp$ interaction vertex. In particular, the impact parameter requirements suppress the heavy flavour hadron content to a negligible level.

The probability of an unstable hadron of mass \(m\), lifetime \(\tau\), and momentum \(p\) to decay within a detector of length $d$ is
\begin{equation}
  1 - \exp\left(-\frac{m d}{\tau p}\right) \approx \frac{m d}{\tau p}.
  \label{eq:qcd_bgd:decay} 
\end{equation}
Similar kinematic distributions are predicted for pions, kaons and protons in the simulation.
Therefore, the in-flight decay background can be modelled by the data with weights of $1/p$.
The majority of the in-flight decays occur outside the magnetic field region and therefore have minimal influence on the measured momentum.
The absolute normalisation is not needed because this background component is allowed to vary freely in the $m_W$ fit.

The weighted \pt spectra for both charges are shown in Fig.~\ref{fig:qcd_bgd_fit} and are overlaid
with best-fit functions of the form~\cite{Hagedorn:1983wk},
\begin{equation}
  \left(1 + \frac{\pt}{a}\right)^{-n},
  \label{eq:qcd_bgd:hagedorn}
\end{equation}
where $a$ and $n$ are empirical parameters that are determined in the fits.
In addition to giving a good fit to the data, this functional form is verified to describe the pion and kaon spectra in simulation.
The fit functions are sampled to generate background candidates for inclusion in the $m_W$ fit.
A charge asymmetry of $\mathcal{O}$(10\%), favouring positively charged hadrons due to the $pp$ initial state, is observed and is included in the sampling.

The uncertainty in the hadronic background model is dominated by systematic sources.
Three different systematic uncertainties, which are combined in quadrature, are assessed.
The first accounts for assuming that the data sample can be treated as containing a single hadron species.
The second accounts for a small bias from the inverted muon identification requirements.
The third accounts for a small dependence on the range of $\pt$ values used in the fits to Eq.~\ref{eq:qcd_bgd:hagedorn}.
The combined systematic uncertainty is $2\mev$.

\section{Modelling \boldmath{\PW} and \boldmath{\PZ} boson production}
\label{sec:Physics}

The emulation of different $m_W$ hypotheses is achieved by assigning event weights based on a relativistic Breit-Wigner function with a mass-dependent width.
Further weight-based corrections are also applied to simulated \PW and \PZ boson events to improve upon the limited formal accuracy of \pythia.
A weighting to model QCD effects is applied in the basis of Eq.~\ref{eq:DiffXsec}.
A further weighting to model QED effects is applied as a function of the logarithm of the relative energy difference between the dimuon system before and after QED final-state radiation.

Higher order electroweak corrections are not included in the model.
Instead, an uncertainty of 5\mev is attributed to these missing corrections
using samples of events generated using \powhegbox~\cite{POWHEGewW,POWHEGewZ,EWCorrections} with and without electroweak corrections.
These events are interfaced with \pythia and the uncertainty is evaluated
using the {\em data challenge} methods described in Sect.~\ref{subsec:data_challenges}.

\subsection{Candidate QCD programs}

Five software programs, or combinations of these, are evaluated as potential candidates for the weighting of the simulated \PW and \PZ boson events.
\begin{enumerate}
\item \pythia: Events are generated using \pythia version 8.235~\cite{Pythia8Main} with several values of the intrinsic transverse momentum (\IKT) of the initial state partons and $\alpha_s$, closely following the work of Ref.~\cite{Lupton}.
  The \texttt{NNPDF23\_lo\_as\_0130\_qed}~\cite{Ball:2013hta} PDFs are used in the event generation.
  The events are weighted using the methods described in Ref.~\cite{lhapdf6} to the \texttt{NNPDF31\_lo\_as\_0118}~\cite{Ball:2017nwa} and CT09MCS~\cite{Lai:2009ne} PDFs.
\item \powhegpythia: Events are generated using \powhegbox~\cite{POWHEGBoxV2SingleBoson} with the \texttt{NNPDF31\_nlo\_as\_0118} PDFs
  and are subsequently showered with \pythia version 8.244~\cite{Pythia8Main}. The event generation with \powhegbox is repeated with different values of $\alpha_s$. The default Monash~\cite{MonashTune} tune of \pythia is used but event samples are generated with different values of \IKT, and with the same value of $\alpha_s$ as used in \powhegbox.
  This results in a grid of predictions with different $\alpha_s$ and \IKT values. 
\item \powhegherwig: Events are generated equivalently to those from \powhegpythia but substituting \pythia with \herwig~\cite{Bellm:2015jjp} for the parton-shower stage.
\item \herwig: These events are also equivalent to those of \powhegpythia except that the hard process and the parton shower are both
  fully implemented in \herwig~\cite{Bellm:2015jjp}. 
\item \dyturbo: The cross-sections and angular coefficients are computed at $\mathcal{O}(\alpha_s^2)$ accuracy using \dyturbo~\cite{DYTurbo2020} with the \texttt{NNPDF31\_nnlo\_as\_0118} PDFs~\cite{Ball:2017nwa}.
  Predictions for the unpolarised cross-section include resummation to next-to-next-to-leading logarithms and are produced with several values of the $g$ parameter that controls nonperturbative effects.
\end{enumerate} 
Histograms of the unpolarised cross-section in Eq.~\ref{eq:DiffXsec} and the angular coefficients are produced for all combinations of programs and tuning parameters.
These histograms, which are used to determine event weights, have intervals in the transverse momentum, rapidity and mass of the vector boson. %$\pt^V$, $y$ and $M$.

\subsection{QCD weighting and transverse momentum model}

The simulated samples described in Sect.~\ref{sec:Samples} can be weighted in the full five-dimensional phase space of vector boson decays, according to Eq.~\ref{eq:DiffXsec}, to provide predictions based on different models of QCD. For the unpolarised cross-section such weights are found by interpolating between the generated histograms described above.

A detailed measurement of the angular coefficients in $pp\to Z\to\mu\mu$ at $\sqrt{s}=8$~TeV was reported by ATLAS~\cite{Aad:2016izn}.
Predictions based on parton showers are generally found to be unreliable in predicting the angular coefficients.
However, the ATLAS data are reasonably well described by $\mathcal{O}(\alpha_s^2)$ predictions from \textsc{DYNNLO}~\cite{Catani:2009sm}, on which \dyturbo is based.
An exception is the difference between $A_0$ and $A_2$, for which $\mathcal{O}(\alpha_s^2)$ is effectively only leading order, but the present measurement
has a negligible sensitivity to this particular detail.
Hereafter \dyturbo is used in the modelling of the angular coefficients.

Since the prediction of each angular coefficient relies on separate numerator and denominator calculations,
there are four independent renormalisation and factorisation scales that are varied to assess the uncertainty associated with missing higher orders in $\alpha_s$.
In Ref.~\cite{Gauld:2017tww} it is argued that fully correlating the scale variations between the numerator and denominator,
which leads to a large degree of cancellation, may result in inadequate uncertainty coverage.
The present analysis therefore follows the recommendation of Ref.~\cite{Gauld:2017tww}, which is to
vary the four scales independently by factors of $\frac{1}{2}$ and 2
with the constraint that all ratios that could be constructed from the four scales are between $\frac{1}{2}$ and $2$.
This results in an envelope of 31 values of $m_W$ that sets the associated uncertainty. 

\begin{figure}\centering
\includegraphics[width=\DefaultFigWidth]{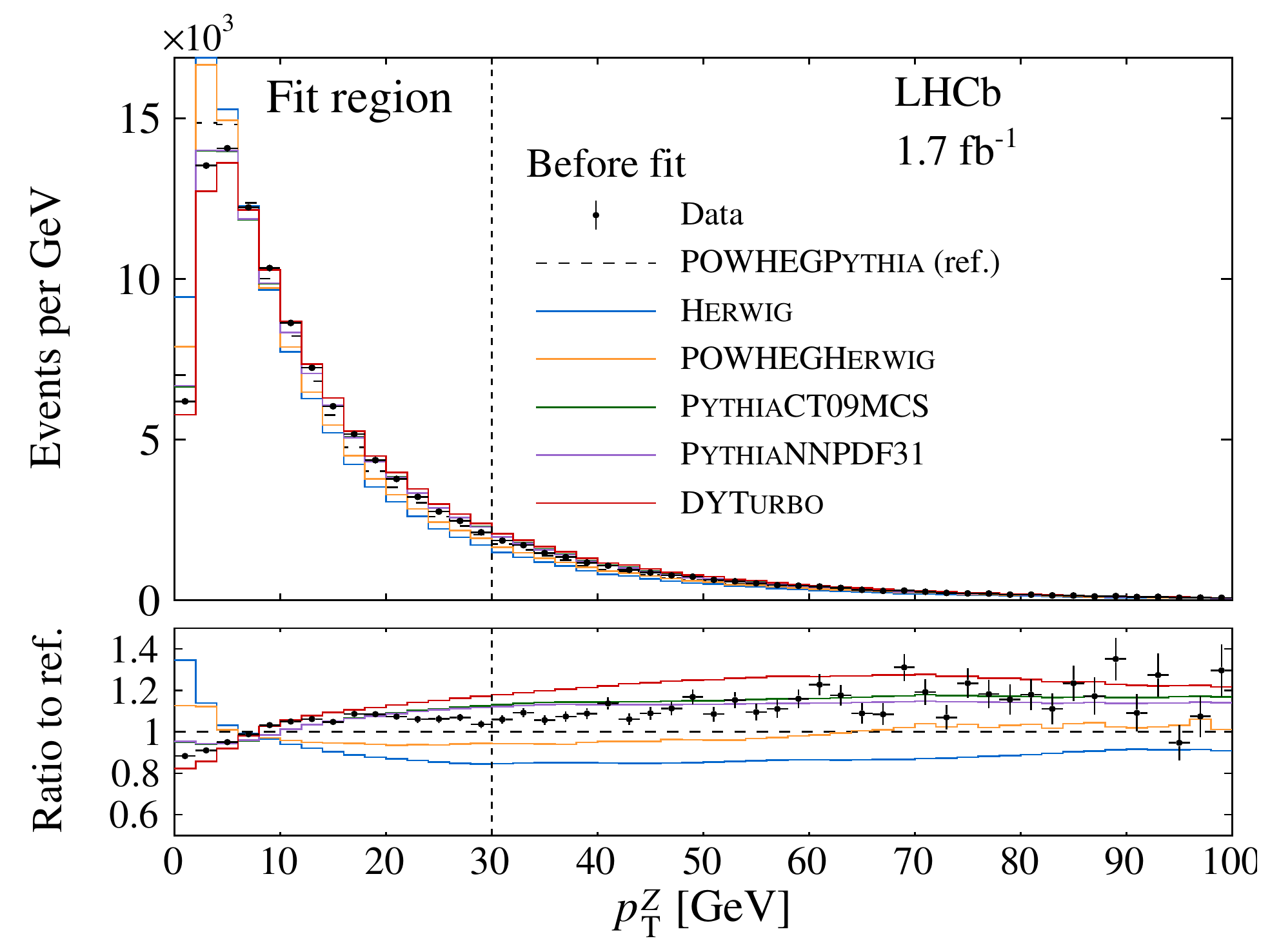}
\includegraphics[width=\DefaultFigWidth]{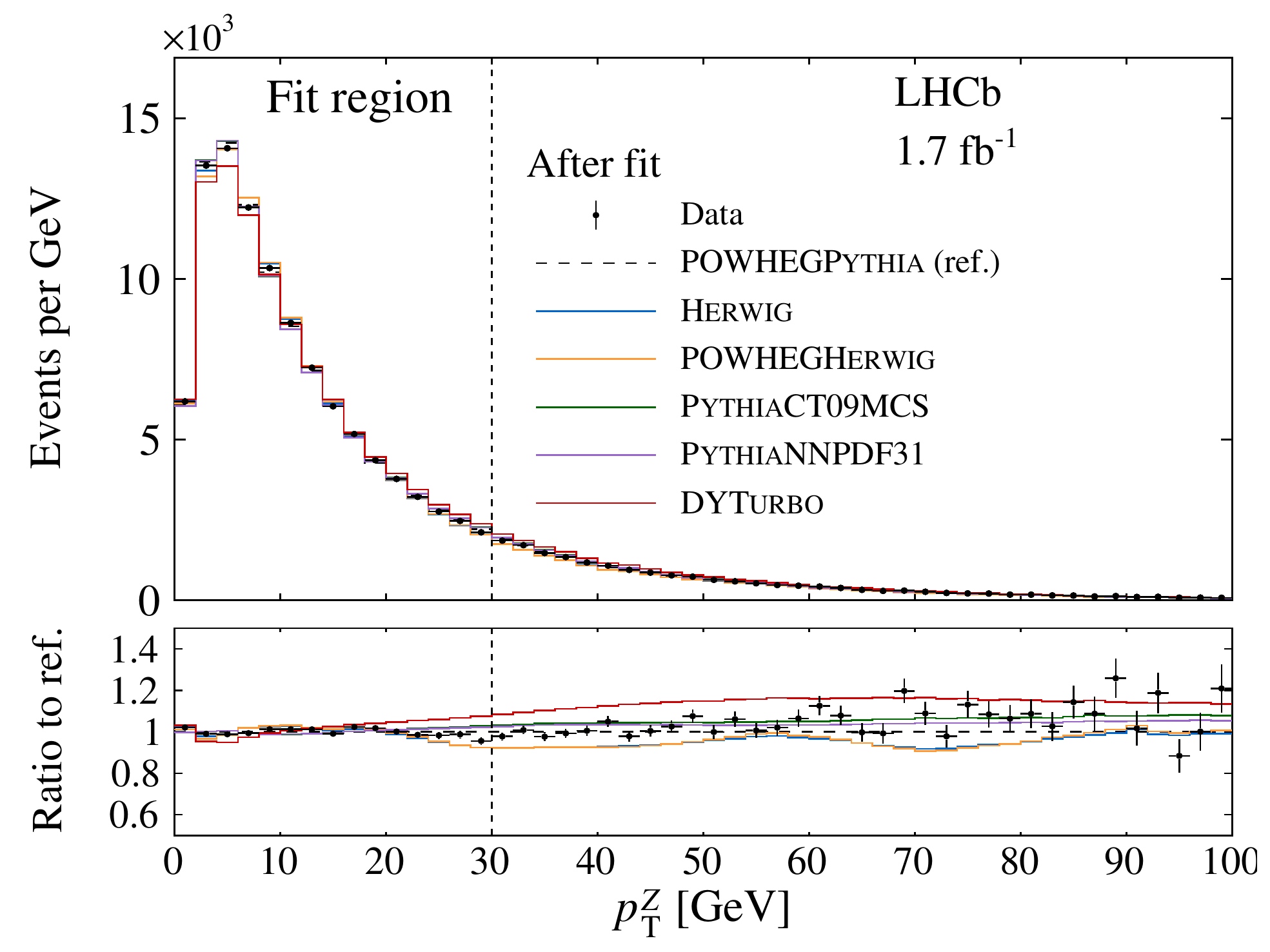}
\caption{\label{fig:ModelValidationPlots}Distributions of $\pt^Z$ (left) before and (right) after the fit for the different
  candidate models of the unpolarised cross-sections. The fit only considers the region $\pt^Z < 30\gev$, indicated by the dashed vertical line. In the lower panels the ratios
with respect to the \powhegpythia model are shown.}
\end{figure}

Figure~\ref{fig:ModelValidationPlots} compares the $\pt^Z$ distribution in the data with the \pythia simulation weighted to the different unpolarised cross-section predictions before and after tuning them to the data.
Table~\ref{tab:validation} lists the $\chi^2$ and the preferred parameter values for the fits with each model.
\dyturbo gives a reasonable prediction but overestimates the number of events with large $\pt^Z$ even with tuning of the $g$ parameter. 
A reasonable initial description is provided by \pythia, which is to be expected since it has already been tuned to $\pt^Z$ data.
The \powhegpythia, \powhegherwig and \herwignlo predictions poorly describe the shape of the $\pt^Z$ distribution with their default values of $\alpha_s = 0.118$.
Their descriptions of the $\pt^Z$ distribution are greatly improved when their $\alpha_s$ and \IKT parameters are tuned.
Of these programs, \powhegpythia gives the most reliable description of the data, with a preferred $\alpha_s$ value of around $0.125$. Large values of $\alpha_s$ are also favoured by other models and in other studies of the $\pt^Z$ distribution~\cite{Ball:2018iqk}.\footnote{The spread in $\alpha_s$ values between the different models also means that this fit result should not be interpreted as a precise and accurate determination of the value of $\alpha_s$.} 
Therefore, \powhegpythia with freely varying $\alpha_s$ and \IKT values is selected for the default fitting model. The systematic uncertainty in the description of the $\pt^Z$ and $\pt^W$ shapes is evaluated with alternative predictions from: \pythia with the  CT09MCS and leading-order NNPDF31 PDFs; \herwignlo and \powhegherwig with the next-to-leading-order NNPDF31 PDFs.
The envelope of shifts in $m_W$ obtained from using these alternative descriptions is found to be $11\mev$, providing the dominant contribution to the systematic uncertainty associated with the modelling of the vector boson transverse momentum.

All of the event generator predictions can be weighted at leading-order to emulate event generation based on different PDFs.
As discussed in Ref.~\cite{lhapdf6} this weighting is not completely valid for events generated at next-to-leading order.
Since \powhegpythia allows {\it in situ} computation of next-to-leading order PDF weights,
it is possible to directly estimate the inaccuracy of the leading-order approximation.
With five \texttt{NNPDF31\_nlo\_as\_0118} replicas it is verified that the differences between the leading-order and next-to-leading-order weighting approaches are smaller than 1\mev in the $m_W$ fit.

Since it is computationally expensive to determine fully the PDF uncertainty in the \dyturbo angular coefficients, the variations in the next-to-leading order PDFs used in the \powhegpythia model of the unpolarised cross-section are
coherently propagated to the angular coefficients.
The \dyturbo angular coefficients are shifted by the differences in values predicted by \powhegpythia in the default (\texttt{NNPDF31\_nlo\_as\_0118}) PDF
compared to the target PDF in the uncertainty assessment.

Separate measurements of $m_W$ based on the NNPDF3.1~\cite{Ball:2017nwa}, CT18~\cite{Hou:2019efy} and MSHT20~\cite{Bailey:2020ooq} PDF sets,
each with their own PDF uncertainty estimate, are reported.
However, since these three sets are based on almost the same data, the central result of this analysis is a simple arithmetic average of the three results, under the assumption that the three PDF uncertainties are fully correlated.

\begin{table}\centering
\caption{\label{tab:validation}Results of fits of different models to the $\pt^Z$ distribution. The uncertainties quoted are statistical, and the $\chi^2$ comparison of the different models to the data is evaluated considering only statistical uncertainties. The right-hand column lists the fit values of the \IKT parameter or, for \dyturbo, the analogous $g$ parameter. The fit with \dyturbo has one more degree of freedom than the fits with the other models since only one tuning parameter ($g$) is used for \dyturbo.
}
\begin{tabular}{lcll}
Program&$\chi^2$/ndf&$\alpha_s$& \\
\hline
\dyturbo &208.1/13&$0.1180$&$g\phantom{^{\rm intr}} = 0.523 \pm 0.047 \gev^2$\\
\powhegpythia &30.3/12&$0.1248 \pm 0.0004$&$\IKT = 1.470 \pm 0.130\gev$\\
\powhegherwig &55.6/12&$0.1361 \pm 0.0001$&$\IKT = 0.802 \pm 0.053\gev$\\
\herwignlo &41.8/12&$0.1352 \pm 0.0002$&$\IKT = 0.753 \pm 0.052\gev$\\
\pythia, CT09MCS&69.0/12&$0.1287 \pm 0.0004$&$\IKT = 2.113 \pm 0.032\gev$\\
\pythia, NNPDF31&62.1/12&$0.1289 \pm 0.0004$&$\IKT = 2.109 \pm 0.032\gev$\\
\hline
\end{tabular}
\end{table}

\subsection{Angular scale factors}

The uncertainties in the angular coefficients from \dyturbo  would lead to an uncertainty of $\mathcal{O}(30)\mev$ in $m_W$, with the dominant contribution attributed to the $A_3$ coefficient. The importance of $A_3$ can be understood by inspection of Eq.~\ref{eq:DiffXsec}: an increase in $A_3$ enhances the cross-section for events with large $\sin\thetaCS$ and $\cos\phiCS$. The contribution to the muon $p_T$ from the $W$ boson mass scales with $\sin\thetaCS$ while the contribution from the transverse momentum of the $W$ boson scales with $\pm\cos\phiCS$ for $W^{\pm}$ boson production.
By allowing a single $A_3$ scaling factor, which is shared between the $W^+$ and $W^-$ processes, to vary freely in the $m_W$ fit the angular coefficient uncertainty is reduced by roughly a factor of three, to $10 \mev$.
Effectively the resulting model only depends on \dyturbo for the kinematic dependence of $A_3$, while all other coefficients are fully modelled by \dyturbo.

\subsection{Parametric correction at high transverse momentum}
\label{sec:kfactor}
\begin{figure}\centering
  \includegraphics[width=\DefaultFigWidth]{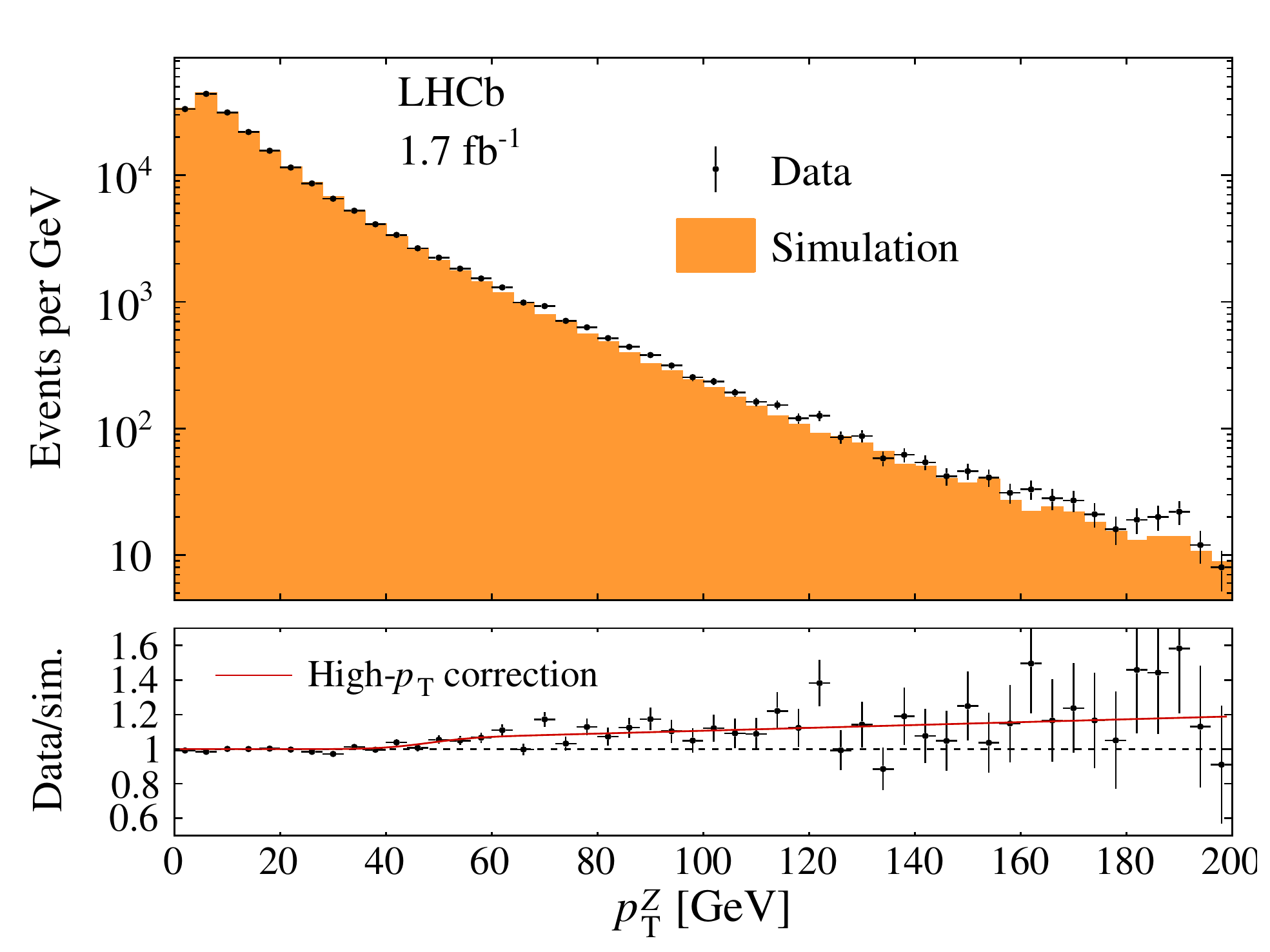}
  \caption{\label{fig:kfactor}Distribution of $\pt^Z$ compared to the \powhegpythia model prior to the parametric correction, which is delineated by the red line in the lower panel.}
\end{figure}

While \powhegpythia is shown in Sect.~\ref{sec:Physics} to describe the $\pt^Z$ distribution in the region below 30\gev,
it systematically underestimates the cross-section at higher $\pt^Z$.
This is expected due to the missing matrix elements for the production of a weak boson and more than one jet.
Figure~\ref{fig:kfactor} compares the $\pt^Z$ distribution in the data with the model prediction having set $\alpha_s$ and \IKT to be close to the final fit values.
For $\pt^Z \geq 40$\gev the model starts to underestimate the cross-section, reaching the ten per cent level at $\pt^Z \sim 100$\gev.
In the lower panel of Fig.~\ref{fig:kfactor} the data to prediction ratio is overlaid with a function of the form
\begin{equation}
  (1 + p_0 + p_0\mathrm{Erf}(p_1(\pt^V-p_2)))\times (1+p_3 \pt^V).
\end{equation}
Since the universality of this correction between \PW and \PZ boson processes is not well controlled,
an uncertainty of 100\% of this correction is included as an additional systematic uncertainty associated with the vector boson \pt model. This contributes an uncertainty in the $m_W$ value smaller than $1\mev$.

\subsection{QED weighting} 

\begin{figure}
    \centering
    \includegraphics[width=\DefaultFigWidth]{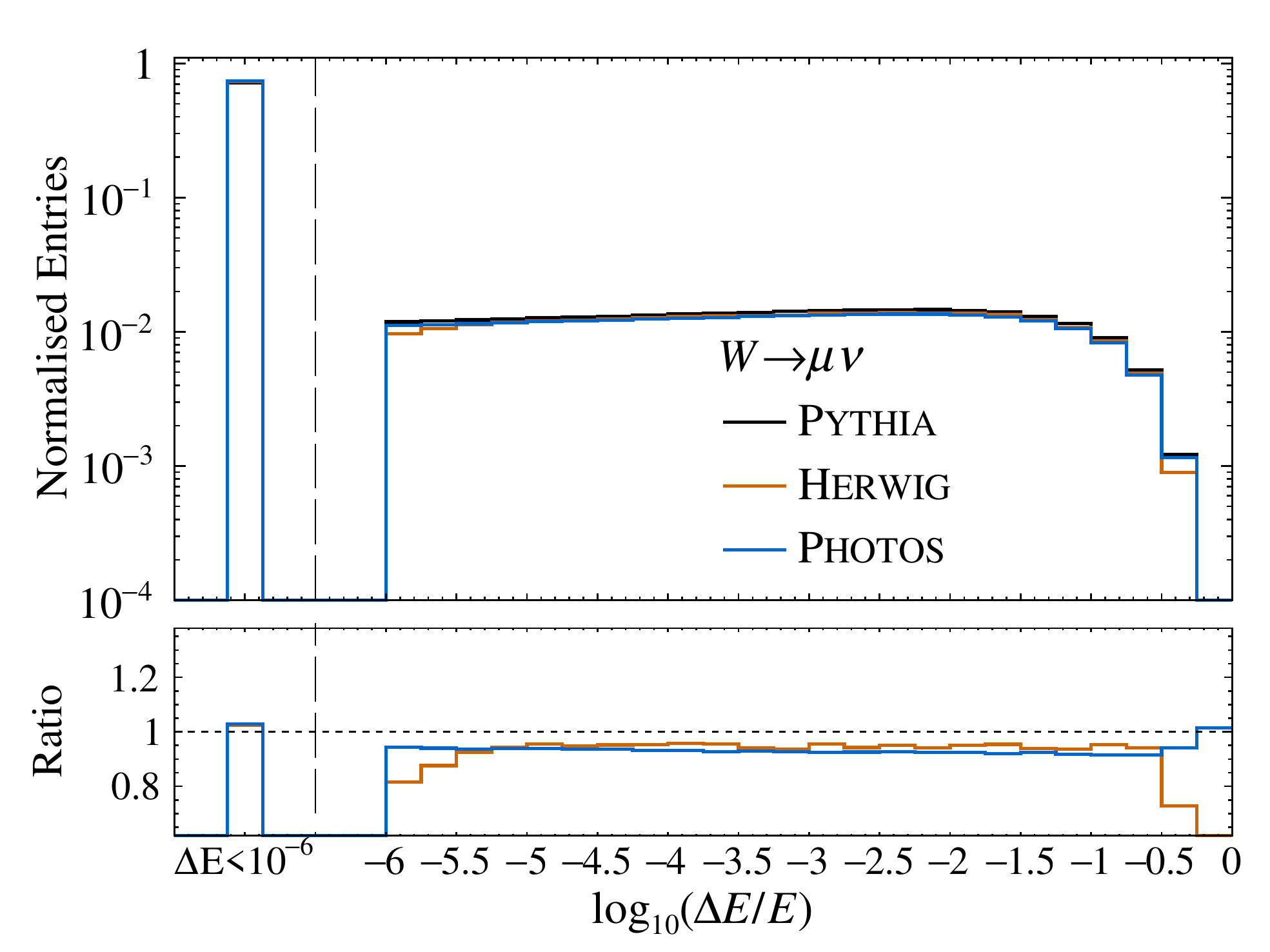}
    \includegraphics[width=\DefaultFigWidth]{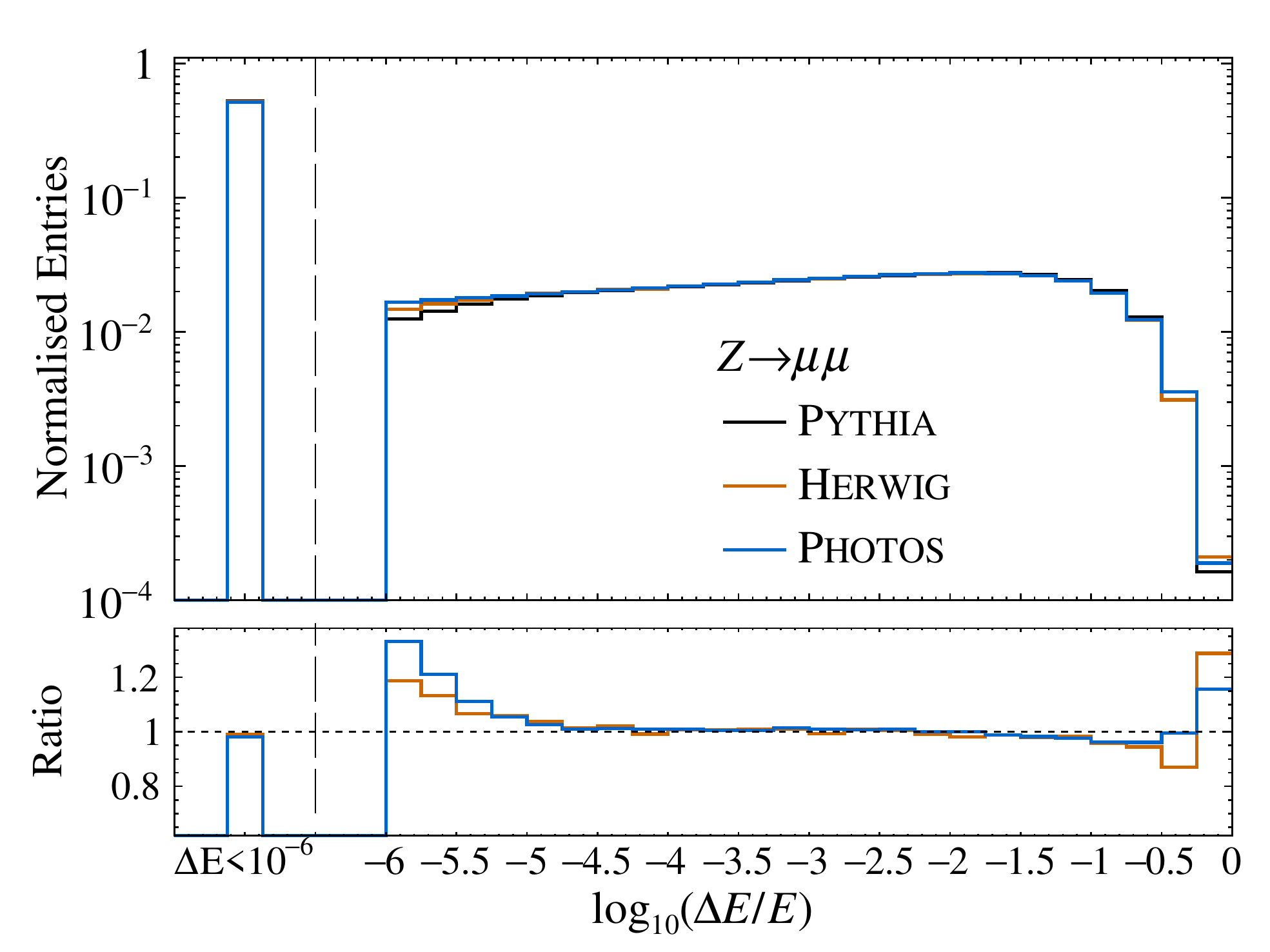}
    \caption{Logarithm of the relative energy loss of the dilepton system due to final-state radiation for (left) \PW boson events and (right) \PZ boson events. An energy loss of below $10^{-6}$ is considered unresolvable and is accounted for in the underflow bin to the left of the dashed vertical line. In the lower panel the ratio with respect to \pythia is shown.}
    \label{fig:qedfsr}
\end{figure}

The effect of the QED final-state radiation is largely characterised by the energy difference between the final-state lepton system before and after radiation.
The logarithm of this energy difference as described by \pythia, \herwig, and an alternative configuration of \pythia with the final-state radiation modelled by \photos~\cite{davidson2015photos} is shown in Fig.~\ref{fig:qedfsr}.
Event-by-event weights are evaluated for the \herwig and \photos models relative to \pythia and applied to the simulated samples used in the analysis.
The default model uses the arithmetic average of the \herwig, \photos and \pythia weights, where the \pythia weights are equal to unity.
The systematic uncertainty, amounting to $7 \mev$, is taken from the envelope of $m_W$ values corresponding to each of the three models taken individually.

\section{\boldmath{\PW} boson mass fit}
\label{sec:MassFit}
The $m_W$ value is determined through a simultaneous fit of the $q/\pt$ distribution of \PW boson candidates
and the $\phi^{\ast}$ distribution of \PZ boson candidates.
The $\phi^{\ast}$ variable is preferred over $\pt^Z$ because it is less susceptible to several of the modelling uncertainties, while still being sensitive to the
parameters affecting the predicted $\pt^Z$  and $\pt^W$  distributions.
The $\phi^{\ast}$ distribution extends to $\phi^{\ast} = 0.5$ while the $q/\pt$ distribution includes two fit regions covering $28 < \pt < 52$\gev.
Projections of the $q/\pt$ distribution cover a wider interval with $\pt > 24$\gev that includes regions outside the fit.
The model of both distributions is based on simulated event samples with event-by-event weights. 
The fit minimises the sum of two negative log-likelihood terms, associated with the $q/\pt$ and $\phi^{\ast}$ distributions,
that are computed using the Beeston-Barlow-Lite prescription~\cite{BeestonBarlow:1993},
which accounts for the finite size of the simulated samples.
The sum of these terms multiplied by a factor of two is denoted as the $\chi^2$.

The $\phi^{\ast}$ distribution is modelled including background contributions from $Z\to\tau\tau$ and top quarks.
The model of the $q/\pt$ distribution includes the dominant 
$W\to\mu\nu$ signal component and several background sources.
The largest background, with a fraction of around $7 \times 10^{-2}$, is attributed to \Zmm,
which is simulated with true invariant masses above 20\gev.
The $W\to\tau\nu$ and hadronic background components each contribute at the $\mathcal{O}(10^{-2})$ level.
A combination of rarer background sources including $Z\to\tau\tau$, top quarks, vector boson pairs, and heavy flavour hadrons
gives a total contribution below $10^{-2}$.

The fractions of the $W^+$ and $W^-$ signal components and the hadronic background are allowed to vary freely.
The $W\to\tau\nu$ component is constrained, using the known $\tau\to\mu\nu\bar{\nu}$ branching fraction~\cite{PDG2020}, relative to the $W\to\mu\nu$ component.
All other component fractions are fixed relative to the observed number of \PZ boson candidates in the $\phi^{\ast}$ distribution using the fiducial cross-sections for the corresponding processes relative to that of \PZ boson production.
The fiducial selections for \PW and \PZ boson processes are the same as used in measurements of the corresponding cross-sections~\cite{LHCb-PAPER-2015-049}.
For all other processes the fiducial regions correspond to the requirement of a single muon in the region $\pt > 20$\gev and $2 < \eta < 4.5$.
The measured cross-section for $\PZ \to \mu\mu$ is used~\cite{LHCb-PAPER-2016-021}.
The cross-sections for the rare background processes are determined using \powheg with the next-to-leading-order NNPDF3.1 PDF sets.

The shapes of background components arising from the decay of electroweak bosons are determined from the same models used to describe the signal component. Systematic variations of the model used to describe \PW boson production therefore also simultaneously provide systematic variations associated with the shapes of background contributions from the decay of electroweak bosons. The uncertainty in $m_W$ from varying the predicted cross-sections for the rarer background within their uncertainties is negligible.

The default physics model is based on the simulated samples set out in Sect.~\ref{sec:Samples}, fully weighted using
a combination of \powhegpythia and \dyturbo for the QCD description and a combination
of \pythia, \photos and \herwig for the QED description.
The fit is configured to determine the following parameters:
\begin{enumerate}
\item the value of $m_W$,
\item the fraction of $W^+$ signal,
\item the fraction of $W^-$ signal,
\item the fraction of QCD background,
\item the value of $\alpha_s$ for the \PZ boson processes ($\alpha_s^Z$),
\item an independent $\alpha_s$ value that is shared for the $W^+$ and $W^-$ signals ($\alpha_s^W$),
\item a shared \IKT value for all \PW and \PZ boson processes,
\item and an $A_3$ scale factor that is shared by the  $W^+$ and $W^-$ signals.
\end{enumerate}

\subsection{Data challenge tests}
\label{subsec:data_challenges}

\begin{figure}\centering
\includegraphics[width=.48\textwidth]{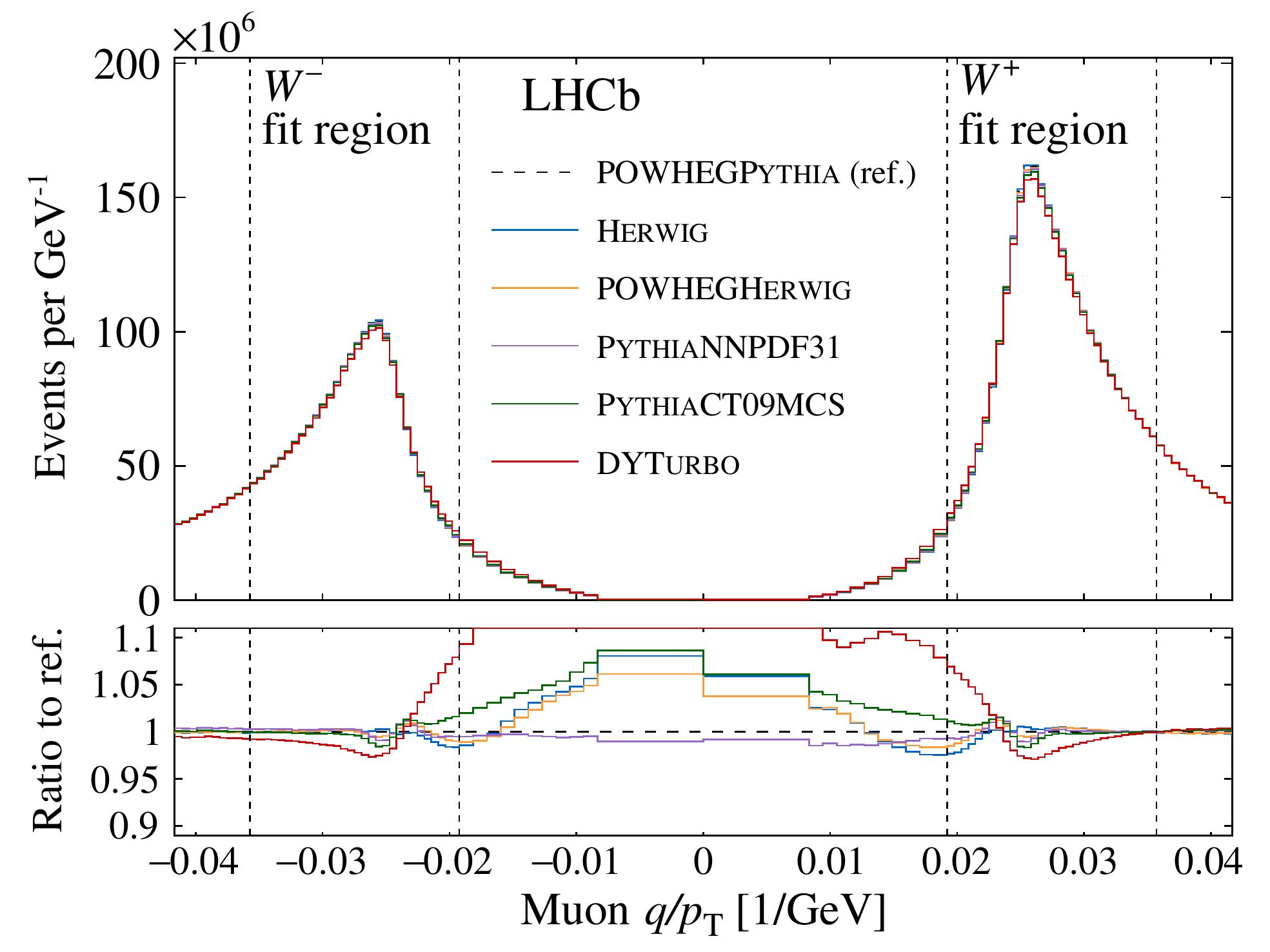}
\includegraphics[width=.48\textwidth]{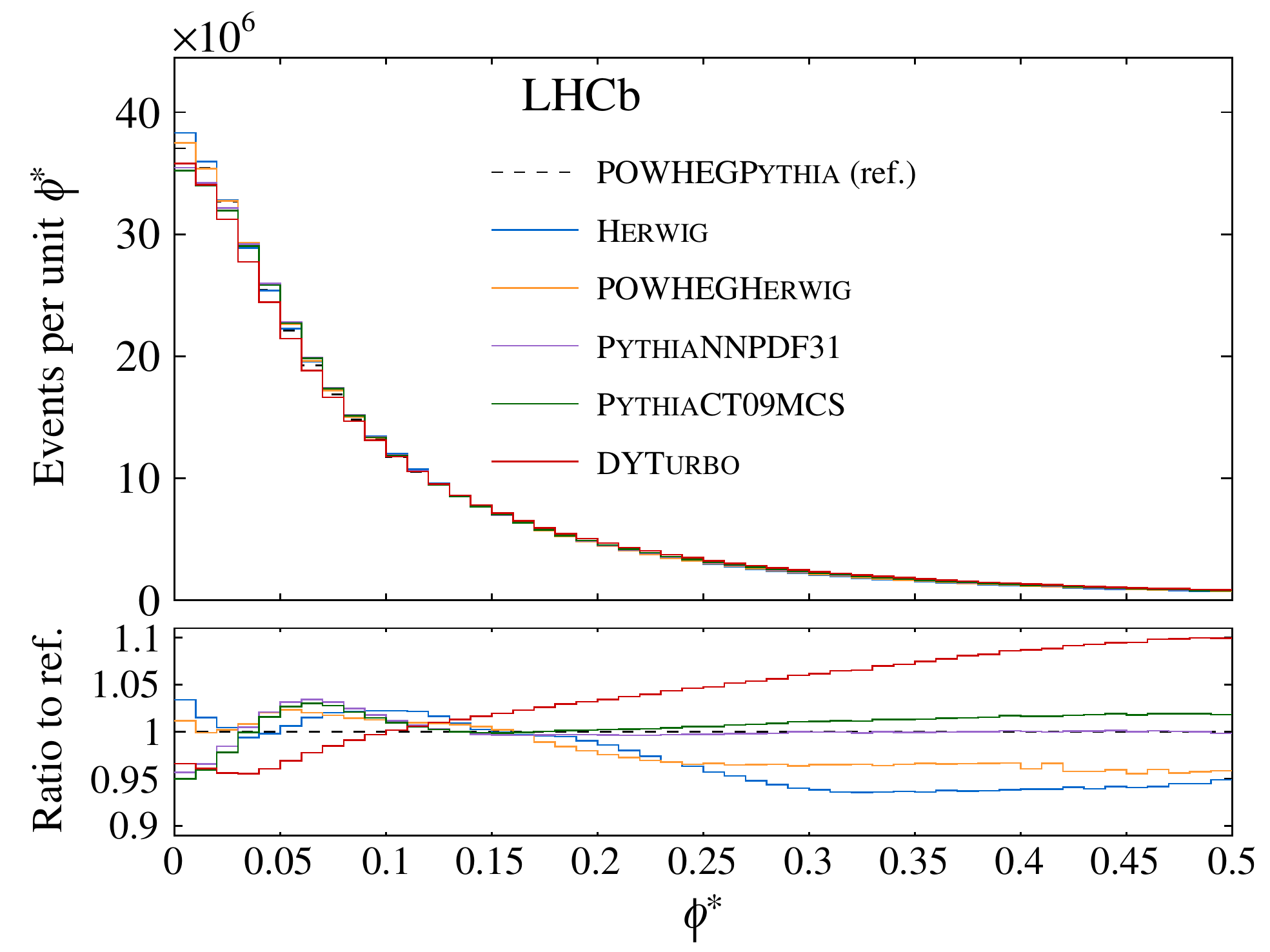}
\caption{\label{fig:data_challenge_summary_projections}Projections of the (left) $q/\pt$ and (right) $\phi^*$ distributions for the challenge datasets. The four dashed vertical lines indicate the two fit regions in the $q/\pt$ distribution.}
\end{figure}

In Sect.~\ref{sec:Physics} it is concluded that \powhegpythia describes the $\pt^Z$ distribution, in the $\pt^Z \leq 30$\gev region, better than the other candidate models.
It is important to demonstrate that the fit can reliably determine $m_W$ if \PW boson production is better described by one or more of the other models.
Several pseudodata samples are prepared in which the underlying \pythia events, without detector simulation, are weighted to match the default \dyturbo and \powhegpythia model
but with the $\pt^V$ distribution modified to match an alternative model.
The $m_W$ fit is configured with a simplified model, without background components, using a statistically independent sample of the same \pythia events without detector simulation.
Figure~\ref{fig:data_challenge_summary_projections} shows the resulting $q/\pt$ and $\phi^*$ distributions of these pseudodata samples.
Variations of up to five per cent are seen in the shape of the $\phi^{\ast}$ distribution.
Within the fit regions in $q/\pt$ variations of several per cent can be seen, while the variations exceed $\mathcal{O}(10^{-1})$ in the high-$\pt$ control region.
However, the fit model is able to absorb these differences in the $\alpha_s$ and \IKT nuisance parameters with variations in the preferred $m_W$ value of no more than 10\mev. Table~\ref{label:DataChallenges} lists the results of the fits to these pseudodata samples.
The observed variation in $m_W$ is consistent with the uncertainty due to modelling the vector boson transverse momentum distribution in the fit to LHCb data, as discussed in Sect.~\ref{sec:Physics}.

\begin{table}\centering
\caption{\label{label:DataChallenges}Fit results with the same default fit model used for the templates
  but with different models used for the pseudodata. The \powhegpythia pseudodata correspond to $\alpha_s=0.125$ and $\IKT=1.8\gev$.
  The \herwignlo and \powhegherwig pseudodata correspond to $\alpha_s=0.136$ and \IKT=1.3\gev.
  The \pythia pseudodata correspond to $\alpha_s=0.127$ and $\IKT=2.7\gev$.
  The \dyturbo pseudodata correspond to $\alpha_s=0.118$ and $g=1\gev^2$.
  The contributions to the total $\chi^2$ from the $q/\pt$ and $\phi^*$ distributions are denoted $\chi^2_W$ and $\chi^2_Z$, respectively.
  The shift in the $m_W$ value with respect to the \powhegpythia pseudodata is denoted $\delta m_W$. The uncertainties quoted are statistical.}
\resizebox{\textwidth}{!}{\begin{tabular}{lrrrccc}
  Data config. & $\chi^2_W$ & $\chi^2_Z$ & $\delta m_W$ [MeV] & $\alpha_s^Z$ & $\alpha_s^W$ & $A_3$ scaling\\
\hline
\powhegpythia & $64.8$ & $34.2$ & -- & $0.1246 \pm 0.0002$ & $0.1245 \pm 0.0003$ & $0.979 \pm 0.029$\\
\herwignlo & $71.9$ & $600.4$ & $1.6$ & $0.1206 \pm 0.0002$ & $0.1218 \pm 0.0003$ & $1.001 \pm 0.029$\\
\powhegherwig & $64.0$ & $118.6$ & $2.7$ & $0.1206 \pm 0.0002$ & $0.1226 \pm 0.0003$ & $0.991 \pm 0.029$\\
\pythia, CT09MCS &$71.0$ & $215.8$ & $-2.4$ & $0.1239 \pm 0.0002$ & $0.1243 \pm 0.0003$ & $0.983 \pm 0.029$\\
\pythia, NNPDF31 &$66.9$ & $156.2$ & $-10.4$ & $0.1225 \pm 0.0002$ & $0.1223 \pm 0.0003$ & $0.967 \pm 0.029$\\
%\dyturbo & 81.5 & 334.3 & -0.8 & $0.1260 \pm 0.0001$ & $0.1276 \pm 0.0003$ & $0.968 \pm 0.029$\\
\dyturbo & $83.0$ & $428.5$ & $4.3$ & $0.1305 \pm 0.0001$ & $0.1321 \pm 0.0003$ & $0.982 \pm 0.028$\\
\hline
\end{tabular}}
\end{table}

\subsection{Fit results}

\begin{figure}\centering
\includegraphics[width=\DefaultFigWidth]{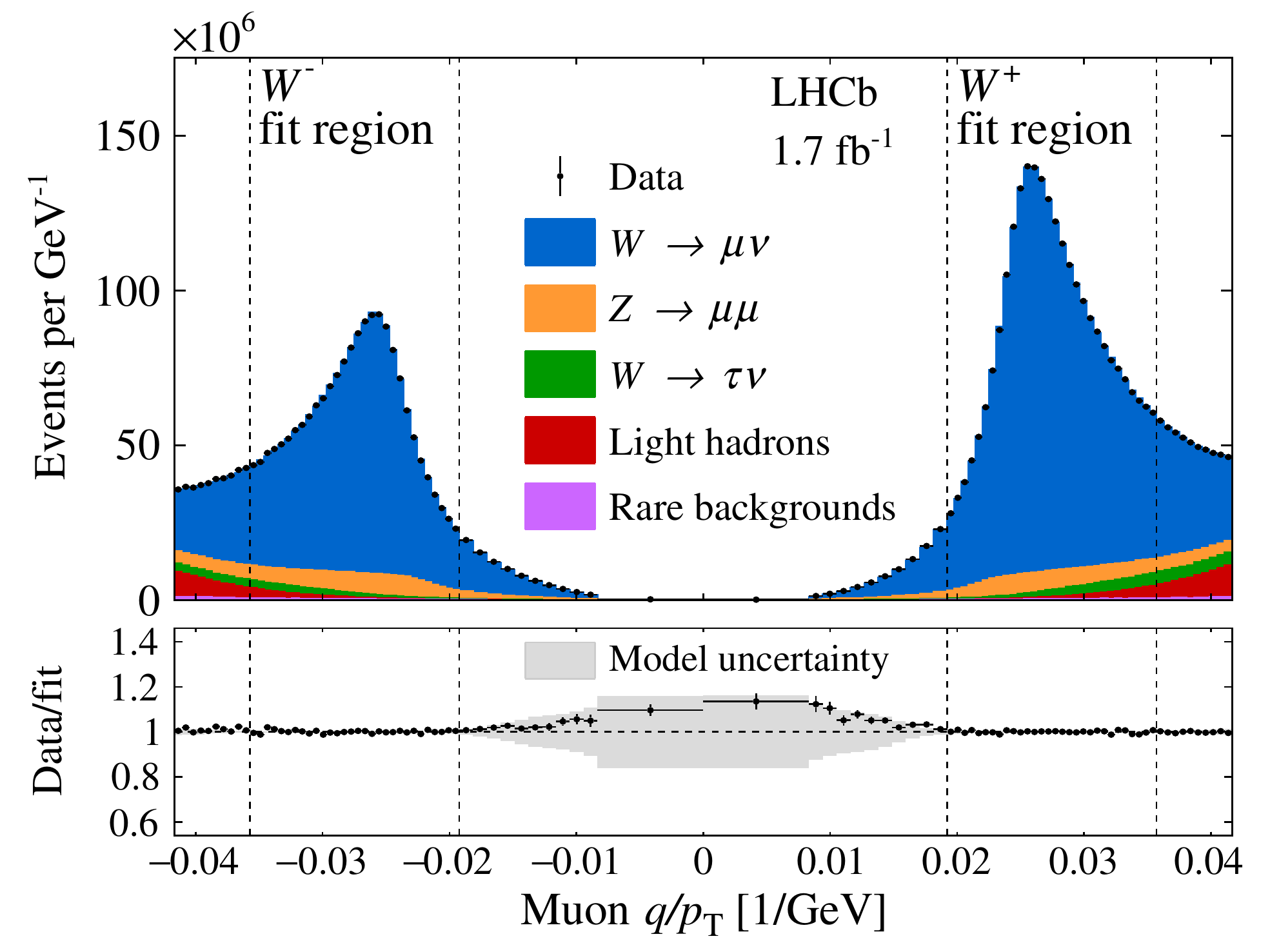}
\includegraphics[width=\DefaultFigWidth]{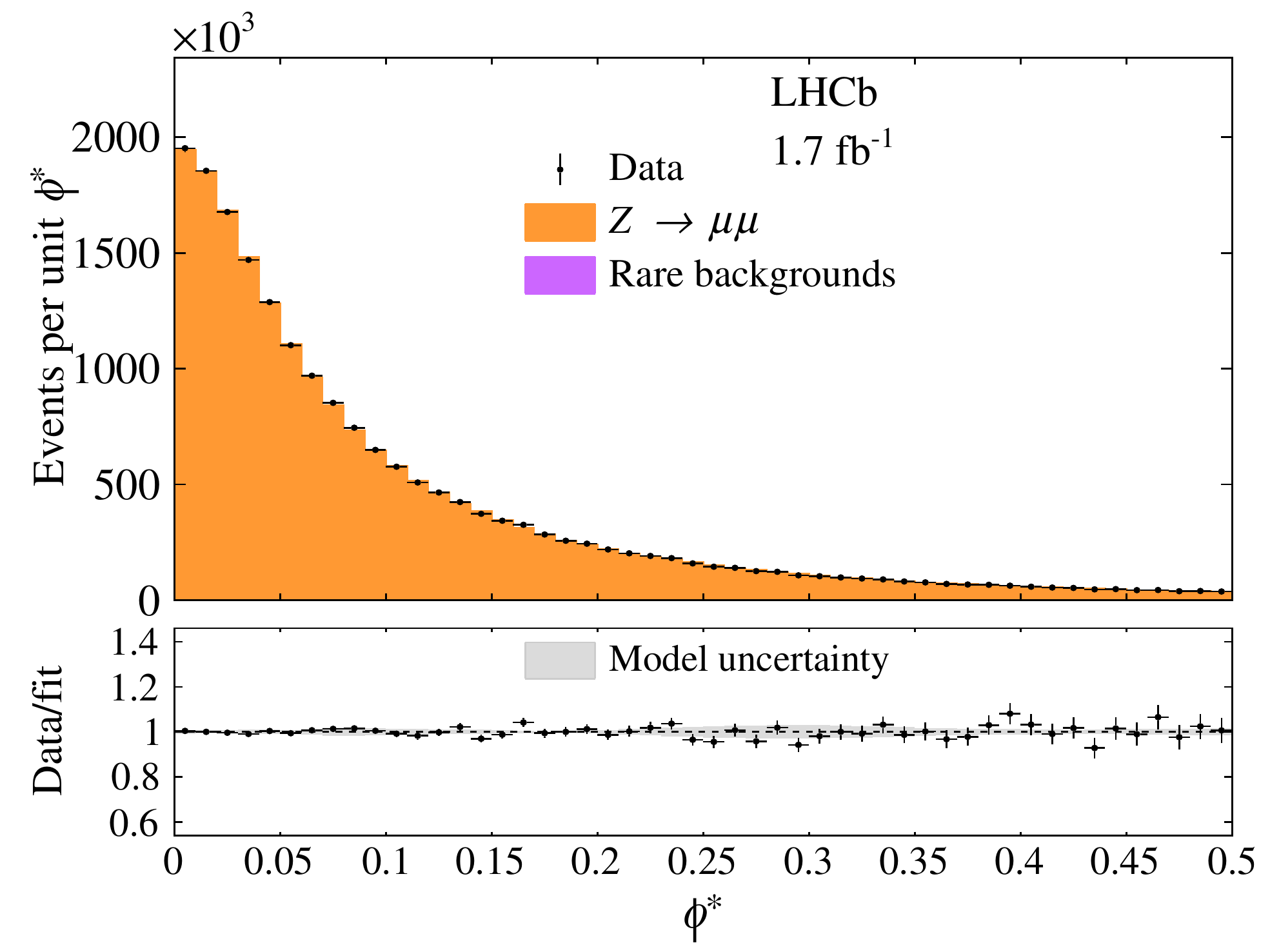}
\caption{\label{fig:remake}Distributions of (left) $q/\pt$ and (right) $\phi^{\ast}$ compared to the model after the $m_W$ fit.}
\end{figure}

The fit to the data, with the \texttt{NNPDF31\_nlo\_as\_0118} PDF set, returns a total $\chi^2$ of 105 for 102 degrees of freedom.
Figure~\ref{fig:remake} compares the $q/\pt$ and $\phi^{\ast}$ distributions from the data with the fit model overlaid.
The model is in good agreement with the data within the fit ranges but it underestimates the high-$\pt$ control region of the $q/\pt$ distribution by up to ten per cent.
This underestimation is within the band of modelling uncertainty, which is dominated by the high-$\pt^V$ parametric correction in that region.
The values of the eight parameters determined from the fit are listed in Table~\ref{tab:results}.
The $\alpha_s$ value for the \PW boson events is roughly $0.002$ higher than for the \PZ boson events.
If the fit is configured with a shared $\alpha_s$ value for the \PW and \PZ boson events the value of $m_W$ changes by $+39\mev$ but the $\chi^2$ is increased by more than 20 units, which strongly favours the configuration with independent $\alpha_s$ values. Furthermore, similar variations between the $\alpha_s$ values for \PW and \PZ boson events are found in the data challenge tests, as shown in Table~\ref{label:DataChallenges}.
The $A_3$ scaling factor is statistically consistent with unity, which suggests that the $\mathcal{O}(\alpha_s^2)$ predictions from \dyturbo, with the central scale choices, are compatible with the data.

\begin{table}\centering
  \caption{\label{tab:results}Values of the parameters determined in the $m_W$ fit with the \texttt{NNPDF31\_nlo\_as\_0118} PDF set. The uncertainties quoted are statistical.}
    \begin{tabular}{l r@{~$\pm$~}l}
    Parameter & \multicolumn{1}{l}{Value} \\
    \hline
    Fraction of $W^+\to\mu^+\nu$ & $0.5288$ & $0.0006$\\ 
    Fraction of $W^-\to\mu^-\nu$ & $0.3508$ & $0.0005$ \\
    Fraction of hadron background & $0.0146$ & $0.0007$ \\
    $\alpha_s^Z$ & $0.1243$ & $0.0004$\\
    $\alpha_s^W$ & $0.1263$ & $0.0003$\\
    \IKT & $1.57$ & $0.14$\gev\\
    $A_3$ scaling & $0.975$ & $0.026$\\
    $m_W$ & \NNPDFmWval & \DataMWStat\mev\\
    \hline
  \end{tabular}
\end{table}

\begin{figure}[!b]\centering
  \includegraphics[width=\DefaultFigWidth]{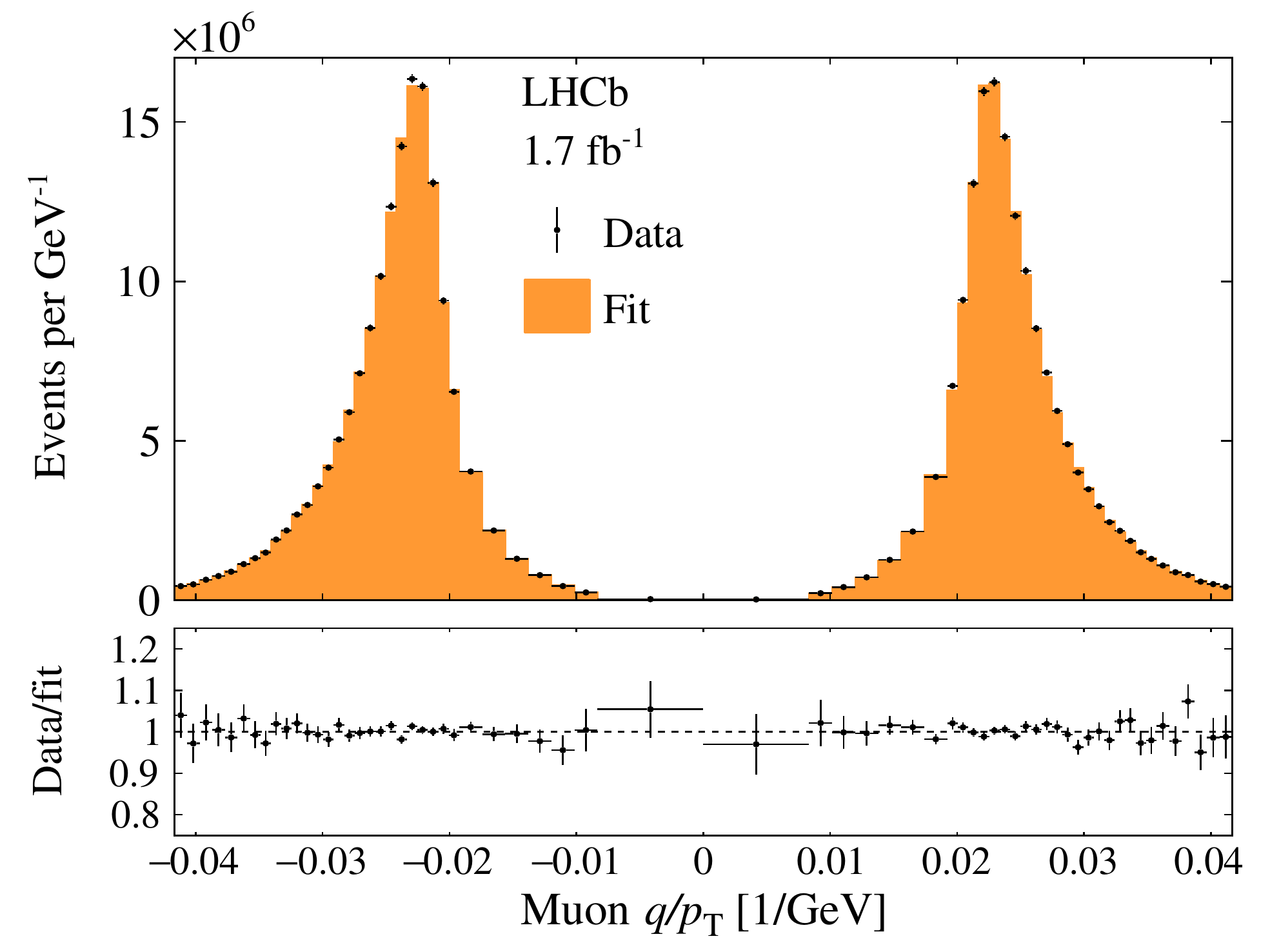}
  \includegraphics[width=\DefaultFigWidth]{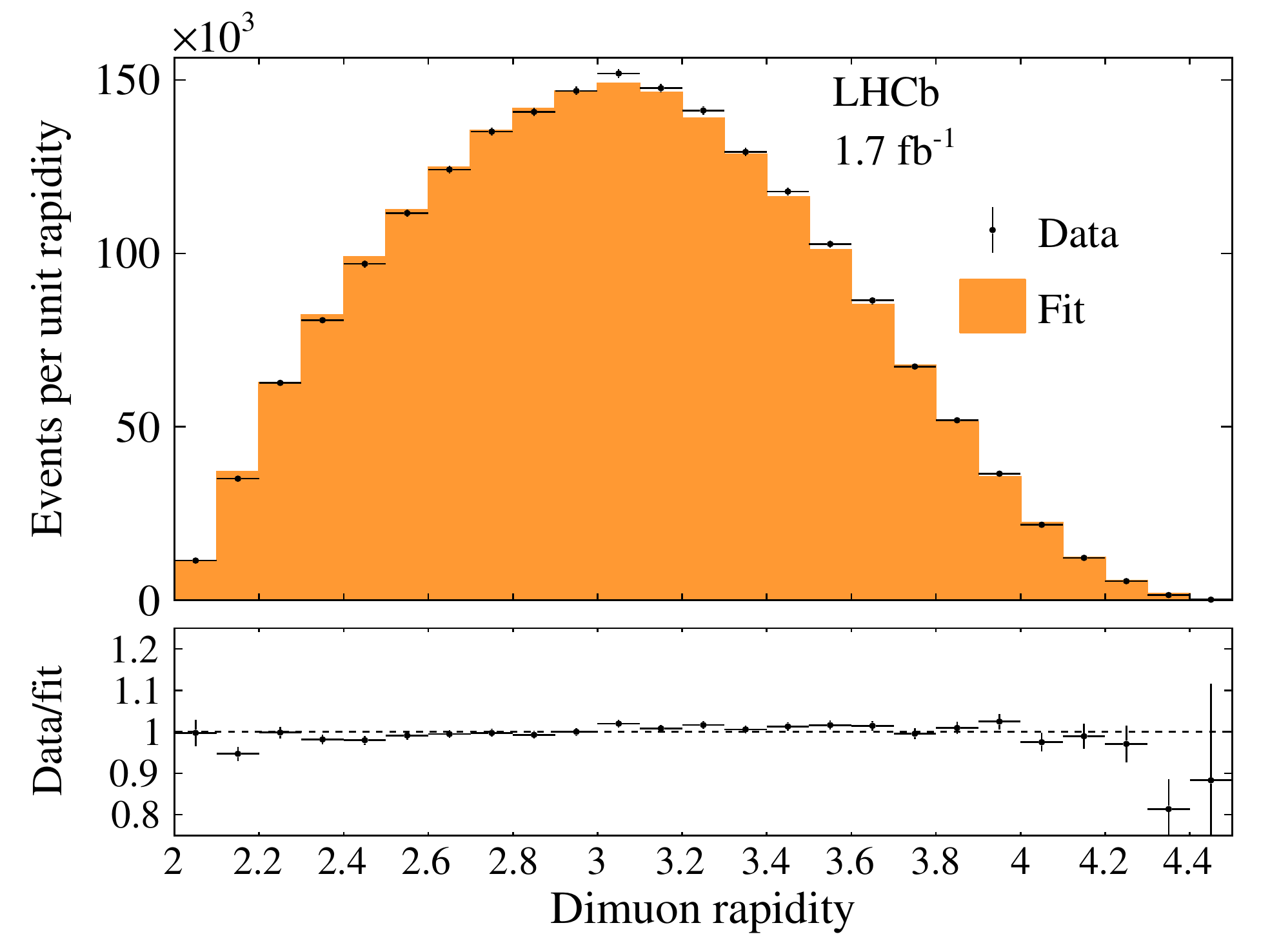}
  \caption[]{Projections of the (left) $q/\pt$ and (right) rapidity distributions for the \PZ boson selection. A final state muon is only included in the $q/\pt$ distribution if it satisfies the \PW boson selection requirements.}
  \label{fig:WMassPostFitDimuon_DEFAULT_13TeV}
\end{figure}

Figure~\ref{fig:WMassPostFitDimuon_DEFAULT_13TeV} (left) shows the projection of
the $q/\pt$ distribution in the \PZ boson sample, where the final state muon is only included if it satisfies the \PW boson selection requirements.
The model is in good agreement with the data.
Figure~\ref{fig:WMassPostFitDimuon_DEFAULT_13TeV} (right) shows that the \PZ boson rapidity distribution is well described by the model.

\section{Systematic uncertainties and cross-checks}
\label{sec:Systematics}
Table~\ref{Tab:pdfs} lists the PDF uncertainties evaluated for fits based on the NNPDF3.1, CT18 and MSHT20 PDF sets.
The $m_W$ values agree within an envelope of $12\mev$, which supports the choice to report an arithmetic average of the three.
With respect to the central value obtained with the NNPDF31 PDFs, the $m_W$ values obtained with the CT18 and MSHT20 PDF sets differ by $-12$~MeV and $-11$~MeV, respectively.
The uncertainties are evaluated according to the specific methods for the three groups.
The NNPDF3.1 uncertainty is evaluated as the RMS of $m_W$ values according to 100 replicas,
whereas the other two sets use fixed numbers of eigenvectors.
The CT18 uncertainty is corrected from 90\% confidence level (CL) to 68\% CL to be consistent with all other uncertainties in this analysis.
For each PDF set, the uncertainty from the replica variations is added in quadrature to the uncertainty from variations in the $\alpha_s$ used in the PDF fits. Values of $0.116 < \alpha_s < 0.120$ are considered, and the uncertainty in $m_W$ is taken as half of the absolute difference between the corresponding shifts in $m_W$~\cite{Butterworth:2015oua}.\footnote{The variations in $\alpha_s$ are larger than the $\pm 0.0015$ variations recommended in Ref.~\cite{Butterworth:2015oua} but this choice avoids further {\it ad hoc} scaling.}
The PDF uncertainty on the arithmetic average of the three results is taken as the arithmetic average of the three uncertainties,
in accordance with the assumption that the uncertainties are fully correlated.

\begin{table}\centering
  \caption{\label{Tab:pdfs}Uncertainties for the NNPDF3.1, CT18 and MSHT20 sets.
    The contributions from the PDF uncertainty with fixed $\alpha_s$ and from the $\alpha_s$ variation are quoted separately as is their sum in quadrature,
    which defines the total uncertainty for each PDF set.}
\begin{tabular}{llll}
Set & $\sigma_{\rm PDF,base}$ [\mev] & $\sigma_{\text{PDF},\alpha_s}$ [\mev] & $\sigma_{\rm PDF}$ [\mev] \\ 
\hline
NNPDF3.1 & $8.3$ & $2.4$ & $8.6$ \\
CT18 & $11.5$ & $1.4$ & $11.6$ \\
MSHT20 & $6.5$ & $2.1$ & $6.8$ \\
\hline
\end{tabular}
\end{table}

Table~\ref{Tab:syst} lists each contribution to the systematic uncertainty in the final result, after averaging results based on the three PDF sets.
The systematic uncertainty is split into three orthogonal components that are combined in quadrature.
The uncertainty due to the description of the parton distribution functions is $\DataMWPDF$\mev.
The remaining theory uncertainty in the modelling of \PW and \PZ boson production is $\DataMWTh$\mev, as described in Sect.~\ref{sec:Physics}, with the largest contribution arising from variations of the transverse momentum model.
The experimental uncertainty is $\DataMWExp$\mev, with the different contributions discussed in Sects.~\ref{section:momentum}, \ref{sec:Efficiencies}, and~\ref{sec:qcd-bgd}.

\newcommand*{\tabindent}{ \hspace{3mm}}
\begin{table}\centering
\caption{\label{Tab:syst}Contributions to the systematic uncertainty in $m_W$. Negligible contributions below $1 \mev$ are not listed.}
\begin{tabular}{ll}
Source & Size [\mev] \\
\hline
Parton distribution functions& $\DataMWPDF$ \\
Theory (excl. PDFs) total & $\DataMWTh$ \\
\tabindent Transverse momentum model & $11$ \\
\tabindent Angular coefficients & $10$ \\
\tabindent QED FSR model & $7$ \\
\tabindent Additional electroweak corrections & $5$ \\
Experimental total & $\DataMWExp$\\
\tabindent Momentum scale and resolution modelling & $7$\\
\tabindent Muon ID, trigger and tracking efficiency & $6$\\
\tabindent Isolation efficiency & $4$\\
\tabindent QCD background & $2$\\
Statistical & $\DataMWStat$\\
Total & $\DataMWTot$\\
\hline
\end{tabular}
\end{table}
%momentum scale is set by quad sum of: Details of the smearing function, Smearing model parameters, Quarkonia QED-FSR 
% Energy Offset Correction renamed to Material Budget

Independently of the systematic uncertainty evaluation, several cross-checks of the measurement are performed.
\begin{itemize}
\item \textbf{Consistency of orthogonal subsets}: The data and simulation are split into orthogonal subsets 
by magnet polarity, the product of the muon charge and polarity, and the $\phi$ and $\eta$ of the muon in the \PW boson selection.
These results are reported in Table~\ref{Tab:xchecks_orthogonal}.
Considering the statistical uncertainties only, all differences are within, or just outside, two standard deviations,
which was predefined as a criterion for this test.
\item \textbf{Fit range}: The minimum and maximum $\pt$ of the fit range in the $q/\pt$ distribution are varied around their default values of $28$\gev and $52$\gev, respectively.
The results are reported in Table~\ref{Tab:xchecks_range}.
Considering the variations in the statistical uncertainties in $m_W$ this test shows that the fit results are stable with respect to variations in the fit range.
\item \textbf{Fit model freedom}: The choice of parameters that are determined in the fit is varied
  and the results are reported in Table~\ref{Tab:xchecks_fit}.
The default fit determines one $\alpha_s$ parameter for the $Z$ processes and a second that is shared between $W^+$ and $W^-$ processes. 
With three $\alpha_s$ parameters there
is only a small change in $m_W$ and the fit quality.
The default fit determines a single floating \IKT parameter that is shared among all three processes. 
Neither the $m_W$ value nor the $\chi^2$ are strongly affected by allowing two (with one shared between the \PWp and \PWm processes) or three \IKT parameters to vary freely.
If the $A_3$ scaling factor is fixed to unity the value of $m_W$ shifts by 7\mev and the $\chi^2$ increases by a few units.
In summary the $m_W$ fit seems to be rather insensitive to all of these variations, except that the data strongly prefer 
independent \powhegpythia tunes for the \PW and \PZ boson production processes.
\item \textbf{Use of NNLO PDF sets}: The PDF set used for the analysis is varied from \texttt{NNPDF31\_nlo\_as\_0118} to \texttt{NNPDF31\_nnlo\_as\_0118}. The shift in $m_W$ is $1\mev$.
\item \textbf{Separate \boldmath{$m_W$} values for \boldmath{\PWp} and \boldmath{\PWm} bosons}: an additional parameter is included in the fit, allowing for separate values of $m_W$ for \PWp and \PWm bosons. This mass difference is found to be consistent with zero within one standard deviation.
\item \textbf{\boldmath{\PW}-like measurement of the \boldmath{\PZ} boson mass:} the same methods are applied to the \PZ boson sample alone, to perform a \PW-like measurement of the \PZ boson mass. The values measured with positive and negatively charged muons agree within one standard deviation and their average is consistent with the PDG average~\cite{PDG2020} within one standard deviation. 
\end{itemize}

\begin{table}\centering
\caption{\label{Tab:xchecks_orthogonal}Fit results where the data and simulation samples are split into two orthogonal subsets. For a given split, the first row is defined as the reference with respect to which the difference in $m_W$, denoted by $\delta m_W$, is defined. The uncertainties quoted on $\delta m_W$ are statistical.}
\begin{tabular}{lll}
Subset & $\chi^2_{\mathrm{tot}}$/ndf & $\delta m_W$ [\mev]\\
\hline
Polarity~$=-1$ & $92.5/102$ & -- \\
Polarity~$=+1$ & $97.3/102$ & $-57.5 \pm 45.4$ \\
$\eta > 3.3$ & $115.4/102$ & --\\
$\eta < 3.3$ & $85.9/102$ & $+56.9 \pm 45.5$ \\
Polarity~$\times$~$q = +1$ & $95.9/102$ & --\\
Polarity~$\times$~$q = -1$ & $98.2/102$ & $+16.1 \pm 45.4$ \\
$|\phi| > \pi/2$ & $98.8/102$ & --\\ % Aside
$|\phi| < \pi/2$ & $115.0/102$ & $+66.7 \pm 45.5$ \\ % C-side
$\phi < 0$ & $91.8/102$ & --\\ %lower
$\phi > 0$ & $103.0/102$ & $-100.5 \pm 45.3$ \\ %upper
\hline
\end{tabular}
\end{table}

\begin{table}\centering
  \caption{\label{Tab:xchecks_range}Fit results with variations in the fit range around the default $\pt^{\rm min} = 28$\gev and $\pt^{\rm min} = 52$\gev.
    The second column lists the $\chi^2$ values, the third column lists the shifts in $m_W$ with respect to the default fit and the third column
  lists the statistical uncertainties in $m_W$.}
\begin{tabular}{llll}
Change to fit range & $\chi^2_{\mathrm{tot}}$/ndf & $\delta m_W$ [\mev] & $\sigma(m_W)$ [\mev]\\
\hline
$\pt^{\rm min} = 24$\gev & $96.5/102$ & $+6.8$ & $19.7$\\
$\pt^{\rm min} = 26$\gev & $97.7/102$ & $+9.6$ & $20.9$\\
$\pt^{\rm min} = 30$\gev & $102.7/102$ & $+3.0$ & $25.7$\\
$\pt^{\rm min} = 32$\gev & $84.9/102$ & $-21.6$ & $30.8$\\
$\pt^{\rm max} = 48$\gev & $105.3/102$ & $-3.8$ & $23.2$\\
$\pt^{\rm max} = 50$\gev & $103.0/102$ & $-2.1$ & $23.0$\\
$\pt^{\rm max} = 54$\gev & $96.3/102$ & $-8.6$ & $22.6$\\
$\pt^{\rm max} = 56$\gev & $103.7/102$ & $-14.3$ & $22.4$\\
\hline
\end{tabular}
\end{table}

\begin{table}\centering
\caption{\label{Tab:xchecks_fit}Fit results with variations in which physics parameters are varying freely.}
\begin{tabular}{llll}
Configuration change & $\chi^2_{\mathrm{tot}}$/ndf & $\delta m_W$ [\mev] & $\sigma(m_W)$ [\mev]\\
\hline
$2\to3$ $\alpha_s$ parameters & $103.4/101$ & $-6.0$ & $\pm 23.1$\\
$2\to1$ $\alpha_s$ and $1\to2$ \IKT parameters & $116.1/102$ & $+13.9$ & $\pm 22.4$\\
$1\to2$ \IKT parameters & $104.0/101$ & $+0.4$ & $\pm 22.7$\\
$1\to3$ \IKT parameters & $102.8/100$ & $-2.7$ & $\pm 22.9$\\
No $A_3$ scaling & $106.0/103$ & $+4.4$ & $\pm 22.2$\\
Varying QCD background asymmetry & $103.8/101$ & $-0.7$ & $\pm 22.7$\\
\hline
\end{tabular}
\end{table}

\begin{figure}[h]
    \centering
    \includegraphics[width=\textwidth]{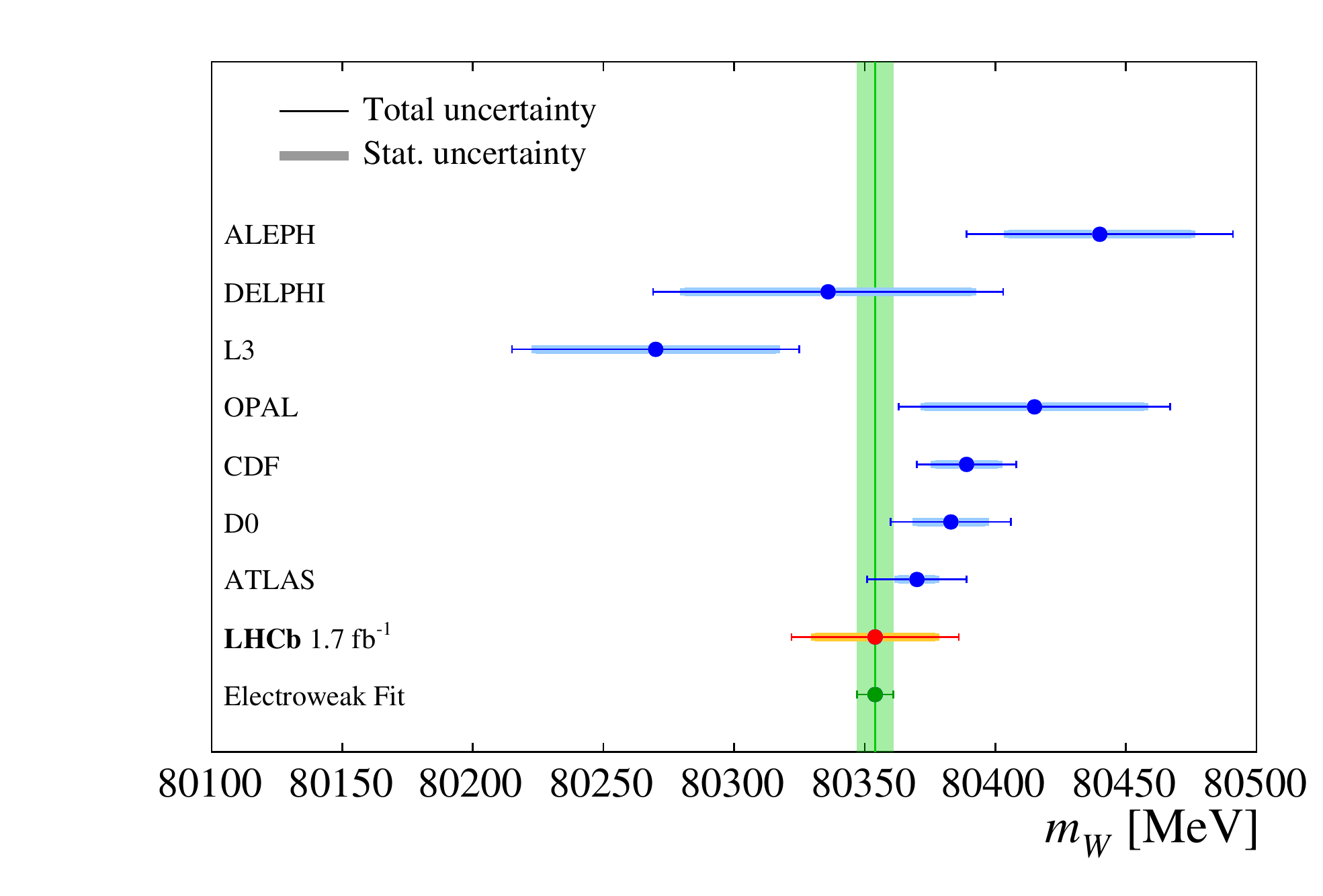}
    \caption{Measured value of $m_W$ compared to those from the ALEPH~\cite{ALEPH:2006cdc}, DELPHI~\cite{DELPHI:2008avl}, L3~\cite{L3:2005fft}, OPAL~\cite{OPAL:2005rdt}, CDF~\cite{CDF}, D0~\cite{D0} and ATLAS~\cite{ATLAS} experiments. The current prediction of $m_W$ from the global electroweak fit is also included.}
    \label{fig:mW_summary_experiments}
\end{figure}

\section{Summary and Conclusion}
\label{sec:Conclusion}

This paper reports the first measurement of $m_W$ with the LHCb experiment.
A data sample of $pp$ collisions at $\sqrt{s}=13 \tev$ corresponding to an integrated luminosity of  1.7~fb$^{-1}$ is analysed.
The measurement is based on the shape of the $\pt$ distribution of muons from \PW boson decays.
A simultaneous fit of the $q/\pt$ distribution of \PW boson decay candidates and of the $\phi^{\ast}$ distribution of \PZ boson decay candidates is verified to reliably determine $m_W$.
This method has reduced sensitivity to the uncertainties in modelling the \PW boson transverse momentum distribution compared to previous determinations of $m_W$ at hadron colliders.
The following results are obtained
\begin{align*}
  m_{W} &= \NNPDFmWval \pm \DataMWStat_{\rm stat} \pm \DataMWExp_{\rm exp} \pm \DataMWTh_{\rm theory} \pm \NNPDFmWerr_{\rm PDF}\mev,\\
   m_{W} &= \CTmWval \pm \DataMWStat_{\rm stat} \pm \DataMWExp_{\rm exp} \pm \DataMWTh_{\rm theory} \pm \CTmWerr_{\rm PDF}\mev,\\
     m_{W} &= \MSHTmWval \pm \DataMWStat_{\rm stat} \pm \DataMWExp_{\rm exp} \pm \DataMWTh_{\rm theory} \pm \MSHTmWerr_{\rm PDF}\mev,
\end{align*}
with the NNPDF3.1, CT18 and MSHT20 PDF sets, respectively. The first uncertainty is statistical, the second is due to experimental systematic uncertainties, and the third and fourth are due to uncertainties in the theoretical modelling and the description of the PDFs, respectively.
Treating the three PDF sets equally results in the following arithmetic average
\begin{equation*}
  m_{W} = \DataMWValue \pm \DataMWStat_{\rm stat} \pm \DataMWExp_{\rm exp} \pm \DataMWTh_{\rm theory} \pm \DataMWPDF_{\rm PDF}\mev.
\end{equation*}
This result agrees with the current PDG average of direct measurements~\cite{PDG2020}
and the indirect prediction from the global EW fit~\cite{Gfitter}, and is compared to previous measurements in Fig.~\ref{fig:mW_summary_experiments}.
This measurement also serves as a first proof-of-principle of a measurement of $m_W$ with the LHCb experiment.
In Ref.~\cite{Pili} it was demonstrated that the PDF uncertainty in a measurement of $m_W$ by LHCb can be strongly reduced by
using {\it in situ} constraints and by fitting the doubly differential distribution of $\pt$ and $\eta$, similar to the measurement by the CMS Collaboration~\cite{CMS:2020cph}, instead of the singly differential $\pt$ distribution.
An approximately three times larger data sample is already available for analysis but particular attention should be paid to reducing the dominant source of systematic uncertainty, which is the modelling of \PW boson production.

% Do not include this in any draft (just for information in the template)
%\input{acknowledgements_intro}
% Comment this in for paper drafts; do not include this in analysis note, conference and figure reports
\section*{Acknowledgements}
%
% These Acknowledgements valid from 3-May-2019
%
\noindent We express our gratitude to our colleagues in the CERN
accelerator departments for the excellent performance of the LHC. We
thank the technical and administrative staff at the LHCb
institutes.
We acknowledge support from CERN and from the national agencies:
CAPES, CNPq, FAPERJ and FINEP (Brazil); 
MOST and NSFC (China); 
CNRS/IN2P3 (France); 
BMBF, DFG and MPG (Germany); 
INFN (Italy); 
NWO (Netherlands); 
MNiSW and NCN (Poland); 
MEN/IFA (Romania); 
MSHE (Russia); 
MICINN (Spain); 
SNSF and SER (Switzerland); 
NASU (Ukraine); 
STFC (United Kingdom); 
DOE NP and NSF (USA).
We acknowledge the computing resources that are provided by CERN, IN2P3
(France), KIT and DESY (Germany), INFN (Italy), SURF (Netherlands),
PIC (Spain), GridPP (United Kingdom), RRCKI and Yandex
LLC (Russia), CSCS (Switzerland), IFIN-HH (Romania), CBPF (Brazil),
PL-GRID (Poland) and NERSC (USA).
We are indebted to the communities behind the multiple open-source
software packages on which we depend.
Individual groups or members have received support from
ARC and ARDC (Australia);
AvH Foundation (Germany);
EPLANET, Marie Sk\l{}odowska-Curie Actions and ERC (European Union);
A*MIDEX, ANR, IPhU and Labex P2IO, and R\'{e}gion Auvergne-Rh\^{o}ne-Alpes (France);
Key Research Program of Frontier Sciences of CAS, CAS PIFI, CAS CCEPP, 
Fundamental Research Funds for the Central Universities, 
and Sci. \& Tech. Program of Guangzhou (China);
%Key Research Program of Frontier Sciences of CAS, CAS PIFI,
%Thousand Talents Program, and Sci. \& Tech. Program of Guangzhou (China);
RFBR, RSF and Yandex LLC (Russia);
GVA, XuntaGal and GENCAT (Spain);
the Leverhulme Trust, the Royal Society
 and UKRI (United Kingdom).

%\input{supplementary}
%\input{appendix}

% This should be taken out in the final paper
%\input{supplementary-app}

\addcontentsline{toc}{section}{References}
%\setboolean{inbibliography}{true}
\bibliographystyle{LHCb}
\bibliography{main,standard,LHCb-PAPER,LHCb-CONF,LHCb-DP,LHCb-TDR}

%\clearpage
%\input{supplementary-app}
%\clearpage
\newpage
% LHCb collaboration author list
% Data extracted on June 9th, 2021 at 11:04am for paper reference LHCb-PAPER-2021-024
\centerline
{\large\bf LHCb collaboration}
\begin
{flushleft}
\small
R.~Aaij$^{32}$,
A.S.W.~Abdelmotteleb$^{56}$,
C.~Abell{\'a}n~Beteta$^{50}$,
T.~Ackernley$^{60}$,
B.~Adeva$^{46}$,
M.~Adinolfi$^{54}$,
H.~Afsharnia$^{9}$,
C.~Agapopoulou$^{13}$,
C.A.~Aidala$^{86}$,
S.~Aiola$^{25}$
% ...

{\footnotesize \it

% ...
$^{9}$Universit{\'e} Clermont Auvergne, CNRS/IN2P3, LPC, Clermont-Ferrand, France\\
% ...
$^{13}$LPNHE, Sorbonne Universit{\'e}, Paris Diderot Sorbonne Paris Cit{\'e}, CNRS/IN2P3, Paris, France\\
% ...
$^{25}$INFN Sezione di Milano, Milano, Italy\\
% ...
$^{32}$Nikhef National Institute for Subatomic Physics, Amsterdam, Netherlands\\
% ...
$^{46}$Instituto Galego de F{\'\i}sica de Altas Enerx{\'\i}as (IGFAE), Universidade de Santiago de Compostela, Santiago de Compostela, Spain\\
% ...
$^{50}$Physik-Institut, Universit{\"a}t Z{\"u}rich, Z{\"u}rich, Switzerland\\
% ...
$^{54}$H.H. Wills Physics Laboratory, University of Bristol, Bristol, United Kingdom\\
% ...
$^{56}$Department of Physics, University of Warwick, Coventry, United Kingdom\\
% ...
$^{60}$Oliver Lodge Laboratory, University of Liverpool, Liverpool, United Kingdom\\
% ...
$^{68}$Syracuse University, Syracuse, NY, United States\\
% ...
$^{86}$University of Michigan, Ann Arbor, United States, associated to $^{68}$\\
\bigskip
% ...
\medskip
}
\end{flushleft}

\end{document}